\documentclass[11pt]{article}

\ifx\pdfoutput\undefined
\usepackage[dvips,bookmarks]{hyperref}
\else
\usepackage{hyperref}
\fi
\hypersetup{colorlinks=false,bookmarksopen,bookmarksnumbered,citecolor=blue,
   pdfstartview=FitH}

\usepackage{amssymb}
\usepackage{makeidx}
\usepackage{graphicx}
\usepackage{ccaption}
\usepackage{subfigure}
\usepackage{rotating}
\usepackage{lscape}
\usepackage{verbatim}
\usepackage{amsmath, amsthm,amssymb}
\usepackage{mathrsfs}
\usepackage{hypernat}
\usepackage{hyperref}
\usepackage{amsfonts}
\usepackage{dsfont}
\usepackage{bbm}
\usepackage{slashed, tensor}

\usepackage{amsopn}

\parskip=\medskipamount

\usepackage[mathscr]{euscript} 

\newcommand{\de}{{\nabla}}             
\newcommand{\deb}{{\overline{\nabla}}}

\usepackage{amsmath}
\usepackage{amsfonts}
\usepackage{slashed}
\usepackage{youngtab}

\usepackage{breqn}
\usepackage{booktabs}

\usepackage{tikz}
\usetikzlibrary{arrows,shapes}

\usepackage{hyperref}

\newcommand {\cC}{{\cal C}}
\newcommand {\cD}{{\cal D}}
\newcommand {\cE}{{\cal E}}
\newcommand {\cF}{{\cal F}}
\newcommand {\cG}{{\cal G}}
\newcommand {\cH}{{\cal H}}

\newcommand {\cK}{{\cal K}}

\newcommand {\cN}{{\cal N}}

\newcommand {\cT}{{\cal T}}
\newcommand {\cU}{{\cal U}}
\newcommand {\cV}{{\cal V}}
\newcommand {\cW}{{\cal W}}
\newcommand {\cX}{{\cal X}}

\def\a{\alpha}
\def\b{\beta}

\def\d{\delta}
\def\e{\epsilon}
\def\f{\phi}
\def\g{\gamma}
\def\G{\Gamma}

\def\l{\lambda}

\def\o{\omega}

\def\q{\theta}

\def\s{\sigma}
\def\t{\tau}

\def\x{\xi}
\def\z{\zeta}
\def\D{\Delta}
\def\F{\Phi}

\def\L{\Lambda}
\def\O{\Omega}

\def\S{\Sigma}

\def\ri{{\rm i}}

\newcommand{\gd}{{\dot\g}}
\newcommand{\dd}{{\dot\d}}
\newcommand{\ad}{{\dot{\alpha}}}
\newcommand{\bd}{{\dot{\beta}}}
\newcommand{\dalpha}{{\dot{\alpha}}}
\newcommand{\dbeta}{{\dot{\beta}}}
\newcommand{\dgamma}{{\dot{\gamma}}}
\newcommand{\ddelta}{{\dot{\delta}}}

\newcommand{\ve}{\varepsilon}
\newcommand{\cDB}{{\overline\cD}}
\newcommand{\DB}{\overline{D}}

\newcommand{\pa}{\partial}
\newcommand{\hf}{\frac12}

\newcommand{\be}{\begin{equation}}
\newcommand{\ee}{\end{equation}}
\newcommand{\bea}{\begin{eqnarray}}
\newcommand{\eea}{\end{eqnarray}}
\newcommand{\non}{\nonumber}
\newcommand{\ba}{\begin{array}}
\newcommand{\ea}{\end{array}}

\newcommand{\bm}[1]{\mbox{\boldmath$#1$}}

\def\double #1{#1{\hbox{\kern-2pt $#1$}}}

\newcommand{\bsubeq}{\begin{subequations}}
\newcommand{\esubeq}{\end{subequations}}

\newcommand{\N}{{\mathcal N}}
\newcommand{\rd}{\mathrm d}
\newcommand{\HC}{{\mathrm{c.c.}}}

\newcommand{\eps}{\varepsilon}

\newcommand{\eol}{\notag \\}
\newcommand{\lc}{{\vert}}

\usepackage{cite}

\newcommand{\loco}{\vert}
\newcommand{\doubar}{{{\loco}\!{\loco}}}

\numberwithin{equation}{section}  

\usepackage{enumitem}

\begin{document}

\begin{titlepage}
\begin{flushright}
May, 2019 \\
\end{flushright}
\vspace{5mm}

\begin{center}
{\Large \bf 
New Fayet--Iliopoulos terms in $\N=2$ supergravity 
}
\\ 
\end{center}
\vspace{2mm}

\begin{center}

{\bf Ignatios Antoniadis${}^{a,b}$, Jean-Pierre Derendinger${}^{a,b}$,
Fotis Farakos${}^{c}$, \\[0.2cm] 
and 
Gabriele Tartaglino-Mazzucchelli${}^{a}$} 
\\
\vspace{10mm}

\footnotesize{
${}^{a}${\it Albert Einstein Center for Fundamental Physics,
Institute for Theoretical Physics,\\
University of Bern,
Sidlerstrasse 5, CH-3012 Bern, Switzerland}}
~\\
\vspace{2mm}
\footnotesize{
${}^{b}${\it Laboratoire de Physique ThÂ\'eorique et Hautes \'Energies - LPTHE\\
Sorbonne Universit\'e, CNRS, 4 Place Jussieu, 75005 Paris, France}  }
~\\
\vspace{2mm}
\footnotesize{
${}^{c}${\it KU Leuven, Institute for Theoretical Physics, \\
			Celestijnenlaan 200D, B-3001 Leuven, Belgium}  }
\vspace{2mm}
~\\
\texttt{
 antoniadis@itp.unibe.ch,
 derendinger@itp.unibe.ch,
fotios.farakos@kuleuven.be,
gtm@itp.unibe.ch}\\
\vspace{2mm}

\end{center}

\begin{abstract}

We present a new type of Fayet--Iliopoulos (FI) terms in $\cN=2$ supergravity that do not 
require the gauging of the $R$-symmetry. 
We elaborate on the impact of such terms on the vacuum structure of the $\N=2$ theory and compare their 
properties with the standard Fayet--Iliopoulos terms that arise from gaugings. 
In particular, we show that, 
with the use of the new FI terms, 
models with a single physical $\cN=2$ vector multiplet can be constructed that give stable de Sitter vacua.

\end{abstract}

\vfill

\vfill
\end{titlepage}

\tableofcontents
\vspace{0.5cm}
\bigskip\hrule

\renewcommand{\thefootnote}{\arabic{footnote}}
\setcounter{footnote}{0}

\allowdisplaybreaks

\section{Introduction}

The vacuum structure of $\cN=2$ supergravity theories has been extensively studied.
Early investigations of $\N=2$ supergravity coupled to abelian vector multiplets \cite{dWLVP, Cremmer:1984hj}
showed already that generating a scalar potential and a breaking pattern requires the {\it gauging} of
some symmetries. Early works considered electric gaugings only, later studies included electric-magnetic 
gaugings and considered the more complicated theories with hypermultiplets, resulting in a variety 
of symmetry and supersymmetry breaking patterns.\footnote{See for instance \cite{DAuria:1990qxt,Andrianopoli:1996vr,Andrianopoli:1996cm,DallAgata:2003sjo,FVP,Trigiante:2016mnt} for reviews. } 
In particular, potentials induced by gauging standard $\N=2$ supergravity coupled to abelian vector multiplets 
arise when Fayet--Iliopoulos (FI) terms \cite{Fayet:1974jb,Fayet:1975yi,VanProeyen:2004xt,Cremmer:1984hj} 
are switched on. These FI terms identify the $R$-symmetries 
of the theory with the symmetries gauged by the $\cN=2$ vector multiplets, 
giving charges and masses to gravitini. 
Moreover, within this setup, 
systems that contain only abelian $\cN=2$ vector multiplets do not admit stable de Sitter vacua. 
For example, with a single physical vector multiplet 
the masses of its two real scalars are restricted by a bound \cite{Catino:2013syn} of the form
\be
\label{noDS}
\text{{\it Standard supergravity with single $\N=2$ vector multiplet}\, :  \quad min}\{m_i^2\} \leq -2 \, {\cal V} \, , 
\ee
which excludes stable de Sitter vacua. 
Similar conditions arise for models with an arbitrary number of abelian vector multiplets, 
leading to the conclusion that the only stable vacua are anti-de Sitter \cite{Cremmer:1984hj}. 
Alternatively, there can be Minkowski backgrounds with flat directions.

To evade the aforementioned restrictions on the vacuum structure of the $\N=2$ supergravity 
one has to drastically deform the theory. 
For example, 
once higher-derivative terms are included in the action, 
the vacuum structure of a supergravity theory is expected to change. 
This direction, for instance, 
has been pursued in the so-called $\cN=1$ {\it pure de Sitter supergravity} 
constructions 
\cite{Antoniadis:2014oya,Hasegawa:2015bza,Antoniadis:2015ala,Bergshoeff:2015tra,Dudas:2015eha,Bandos:2015xnf,Kuzenko:2015yxa,Cribiori:2016qif} 
where higher-derivative terms appear only in the fermionic sector in a controlled non-pathological way
linked to the non-linear realization of supersymmetry. 
Indeed, the constraint \eqref{noDS} arises in $\cN=2$ supergravity
if one restricts the action to contain at most two-derivative terms for both fermions and bosons. 
If this restriction is lifted then new possibilities may arise, 
as can be readily seen from the constructions presented for example in \cite{Kuzenko:2017zla}
that include $\cN=2$ extensions of de Sitter supergravity.

In this work we will present a new deformation in this direction and illustrate its properties. 
In particular, we will investigate the possibility of introducing appropriate 
interactions in the $\N=2$ matter-coupled supergravity 
such that stable de Sitter vacua can be constructed with a minimal number of ingredients. 
We expect our results to have impact to the construction of new general matter-coupled supergravity 
but, for simplicity, in this paper
we will focus on models with a single physical vector multiplet 
and the only new ingredient in our construction will be a new type of Fayet--Iliopoulos term for the $\N=2$ vector multiplet. 
Such deformation has a minimal impact on the bosonic sector of the theory 
and it only affects the scalar potential by introducing an uplifting term. 
Our construction can be considered as the generalization of the new Fayet--Iliopoulos term of $\N=1$ 
supergravity \cite{Cribiori:2017laj,Kuzenko:2018jlz,Antoniadis:2018cpq,Antoniadis:2018oeh} to an $\N=2$ setup. 
The fermionic sector will in principle have a series of higher order terms, 
and will also contain higher order derivative interactions, 
with a structure similar to the non-linear realizations of supersymmetry. 
In the unitary gauge however, where both gravitini are massive, all extra fermionic terms disappear and the Lagrangian simplifies, 
as in the case of $\N=1$ supergravity supplemented with the new FI term.

An essential assumption that enters our construction is that supersymmetry is {\it always} in a spontaneously broken phase. 
The low-energy features of such models have been studied previously in the literature focusing in theories where 
only the $\N=2$ goldstini $\gamma_i^\alpha$ are included in the spectrum 
\cite{Kandelakis:1986bj,KMcA,Cribiori:2016hdz,Kuzenko:2017zla} 
irrespective of the source of the breaking 
($i,j=1,2$ are the $SU(2)_R$ $R$-symmetry indices). 
In our setup, 
once we assume that an $\N=2$ vector multiplet sources the complete $\N=2$ breaking, 
the degrees of freedom of the goldstini will be described by the gaugini $\lambda^i_\alpha$. 
Moreover, 
the auxiliary fields of the vector multiplet $X^{ij}$ will have a non-vanishing vacuum expectation value. 
As we will show, 
we can then consistently construct composite $\N=2$ goldstini of the form 
\be
\gamma_i = -4  \frac{\lambda^j X_{ij}}{X^{pq} X_{pq}} + \dots 
\ee 
Using these composite goldstini we can 
utilize a construction reminiscent of the {\it non-linear realization} of supersymmetry 
and introduce explicit $\N=2$ Fayet--Iliopoulos type of terms that have the form 
\be
{\cal L}_{new~FI} \sim \tilde \xi^{ij} \, X_{ij} + {\cal O}(\gamma\gamma) \, . 
\ee
These new Fayet--Iliopoulos terms will in turn justify 
the initial assumption of spontaneous supersymmetry breaking 
and render the construction self-consistent. 
Indeed, once the 
auxiliary fields are solved by their equations of motion, we find that they receive a non-vanishing vacuum expectation value (vev)
given by $\langle  X_{ij}\rangle \sim \tilde \xi_{ij}$. 
As a result we will see that for the new type of $\N=2$ Fayet--Iliopoulos terms in $\cN=2$ supergravity
the condition \eqref{noDS} 
breaks down and stable de Sitter vacua can be constructed. 
In contrast however to the pure non-linear realizations of supersymmetry 
\cite{Kandelakis:1986bj,KMcA,Cribiori:2016hdz,Kuzenko:2017zla}, 
where only the goldstini appear, 
the component fields in our construction still reside into {\it standard} $\N=2$ supermultiplets.

This article is organized as follows: 
In the next section we review the properties of the goldstini multiplets in global $\N=2$ supersymmetry, 
describe the construction of the {\it new} Fayet--Iliopoulos term for a single vector multiplet 
and contrast its properties with the {\it standard} Fayet--Iliopoulos term. 
In the third section we review technical aspects of the superconformal formulation of $\cN=2$ supergravity in superspace, 
and we elaborate on the standard FI terms focusing on how they give rise to scalar potentials and gaugings. 
In the fourth section we introduce the new type of FI terms that,
in contrast to the standard FI term,
 do not necessarily require the gauging of the $R$-symmetry in supergravity, 
and we study the vacuum structure for the case of a single physical vector multiplet. 
Within this setup we show how the condition \eqref{noDS} is eventually alleviated because of the new Fayet--Iliopoulos terms. 
We discuss our results, together with comments and outlooks, 
in the fifth section while we present some technical details in the appendices.

\section{Deformations of $\cN=2$ global supersymmetry} 
\label{section-2}

In this section we will present the new Fayet--Iliopoulos terms in an $\cN=2$ supersymmetric setup. 
This section serves mostly as a warm-up for the supergravity discussion which follows.

\subsection{$\N=2$ goldstini in global supersymmetry}
\label{section-goldstini-flat}

When 4D $\N=2$ supersymmetry is spontaneously broken to $\N=0$, the effective theory contains two 
fermionic goldstone modes, the goldstini, 
that we call $\gamma_i^\alpha$. 
The $SU(2)_R$ indices $i,j$ take values $1$ and $2$ and refer to the two supersymmetries. 
These fermions have the supersymmetry transformations 
\be
\label{deltagamma}
\delta \gamma_i^\alpha = \epsilon_i^\alpha 
-2\ri \gamma_j \s^m \overline \epsilon^j \partial_m \gamma_i^\alpha \, . 
\ee
The properties of these fermions and their couplings to other fields can be 
conveniently described in superspace. 
The $\N=2$ superspace is parametrized by the coordinates $z_M=(x_m,\,\q^\a_i,\,\bar\q_\ad^i)$ and covariant derivatives
$D_A=(\pa_a,\,D_\a^i,\,\DB^\ad_i)$ satisfying the algebra
\bea
\{D_\a^i,D_\b^j\}=0 \, , \quad 
\{\DB^{\ad}_i,\DB^{\bd}_j\}=0
\, , \quad 
\{D_\a^i,\DB_{\bd j}\}=
-2 \ri \, \d^i_j (\s^a)_{\a \bd} \pa_a \, . 
\eea
The goldstini can then be described by the lowest components of the 
spinor superfields $\Gamma^i_\alpha$, 
which are defined via the constraints \cite{Kandelakis:1986bj} (see also \cite{KMcA,Cribiori:2016hdz,Kuzenko:2017zla}
for a detailed description of $\cN=2\to\cN=0$ Goldstini multiplets)
\begin{eqnarray}
	\label{NSWrepr} 
	\begin{split} 
		D_\alpha^i \Gamma_{\beta j} & = \epsilon_{\beta\alpha} \, \delta^i_j \, , 
		\\
		\overline D_{i \dot \alpha} \Gamma_{\beta j} & = - 2 \ri \, (\sigma^m)_{\rho \dot \alpha} \, \Gamma^{\rho}_i
		\partial_m \Gamma_{\beta j} \, . 
	\end{split}
\end{eqnarray}
The $\cN=2$ supersymmetry transformations take the form 
\be
\label{susySU2}
\delta {\cal O} = \epsilon_i^\alpha  Q^i_\alpha {\cal O} 
+ \overline \epsilon^j_{\dot \alpha} \overline Q_j^{\dot \alpha} {\cal O} \, , 
\ee 
which means that the lowest component of the $\Gamma^\alpha_i$ superfield 
defined as $\gamma_i^\alpha = \Gamma_i^\alpha|$ 
transforms under supersymmetry as \eqref{deltagamma}.  
Notice that the definition of the $\Gamma_i$ in \eqref{NSWrepr} means that it has mass dimension $[\Gamma_i]=-1/2$, 
but one can always rescale with the supersymmetry breaking scale and give to the physical goldstino mass dimension $3/2$. 
The Lagrangian for an $\N=2$ goldstino that does not interact with other superfields has the form 
\be
\label{goldkin}
{\cal L} = - M^4 \int \rd^8 \theta \, |\Gamma|^8 \, , 
\ee
where the real constant $M$ is identified with the supersymmetry breaking scale and has mass dimension $[M]=1$. 
In \eqref{goldkin} we have made use  of the notations
\be
|\Gamma|^8 = \Gamma^4 \overline \Gamma^4 \, , \quad 
\G^4\equiv\frac{1}{3} \, \Gamma^{ij} \Gamma_{ij} =-\frac{1}{3} \, \Gamma^{\a\b} \Gamma_{\a\b}  \, , \quad 
\overline{\G}^4\equiv\frac{1}{3} \, \overline \Gamma_{ij} \overline \Gamma^{ij} 
=-\frac{1}{3} \, \overline \Gamma_{\ad\bd} \overline \Gamma^{\ad\bd}  \, , 
\ee
where we defined $\Gamma_{ij} \equiv \Gamma_{i }^{\alpha} \Gamma_{\alpha j}=\Gamma_{ji}$ 
and $\Gamma_{\a\b} \equiv  \Gamma_{\a}^i\G_{\b i}=\G_{\b\a}$ 
together with their complex conjugates. 
Once we evaluate the superspace integral of \eqref{goldkin} we find the contribution to the vacuum energy
density, the kinetic terms for the two goldstini, 
and a series of higher order self-interactions, 
viz. 
\be
{\cal L} = - M^4 
- \ri M^4 \gamma_i \sigma^m \partial_m \overline \gamma^i 
+ \ri M^4 \partial_m \gamma_i \sigma^m \overline \gamma^i 
+ {\cal O}(\gamma^4) \, . 
\ee

The goldstino superfield $\Gamma_i$ can be also coupled to other $\N=2$ superfields in various ways keeping manifest the 
spontaneously broken supersymmetry. 
We would like however to focus on a specific coupling that will be relevant to our work later. 
Assume we have a scalar $\N=2$ superfield of the form 
\be
{\cal U} = U 
+ \theta_j^\alpha u_\alpha^j 
+ \overline \theta_{\dot \alpha}^j \overline u^{\dot \alpha}_j 
+ {\cal O}(\theta^2) \, , 
\ee
where $U$ is now a scalar field and $u_\alpha^j$ describes fermions appearing at the lowest orders in $\theta$. 
Notice that ${\cal U}$ can be a composite superfield or it can be a descendant of some other superfield 
on which we have acted upon with superspace derivatives. 
We can then consider the term given by 
\be
\label{genU}
\int \rd^8 \theta \, |\Gamma|^8 \, {\cal U} = U 
- \gamma_j^\alpha \, u_\alpha^j  
- \overline \gamma_{\dot \alpha}^j \,  \overline u_j^{\dot \alpha}  
+ {\cal O}(\gamma^2) \, . 
\ee 
In particular, if ${\cal U}$ is a descendant superfield that describes an auxiliary field in its lowest component 
(that is the scalar $U$ transforms as a derivative), 
then in such case \eqref{genU} will provide a linear term in the auxiliary field $U$ followed by terms 
multiplied by increasing powers of goldstini.
In other words, 
the combination $|\G|^8$ of the goldstini can effectively be considered as a 
{\it covariantized} version of a D-term spurion $\q^4\bar\q^4$ \cite{Girardello:1981wz} 
which once multiplied by an arbitrary superfield $\cU$ picks up its lowest component $U$ upon integration over the full superspace. 
A spurion-type F-term, that covariantizes $\q^4$, can also be constructed by simply considering $\overline D^4|\G|^8$.
The difference in using the goldstini instead of the spurions is clearly given by the extra fermionic terms that turn explicit susy breaking 
terms into terms that have the spontenously broken supersymmetry non-linearly realized.
As a result, 
with the use of the goldstino spinor superfield $\Gamma_i^\alpha$ one can always introduce 
in an action supersymmetric terms linear in the scalar auxiliary fields 
of any supermultiplet as in \eqref{genU}. 
This observation is the underlying mechanism utilized in $\N=1$ supergravity to construct a new type of Fayet--Iliopoulos 
term in \cite{Cribiori:2017laj}, 
and we will extend it here to the case of an $\N=2$ vector multiplet. 
To this end, 
we will follow a procedure that requires two steps: 
\begin{enumerate} 
\item Firstly, we will need to construct a composite goldstino spinor superfield $\Gamma_i^\alpha$ 
in terms of the $\N=2$ vector multiplet, 
assuming always that the latter completely breaks supersymmetry to $\cN=0$. 
\item Secondly, 
we will use the composite $\Gamma_i^\alpha$ to construct terms of the form \eqref{genU} that will provide the linear 
terms for the auxiliary fields of the vector multiplet such that supersymmetry is indeed spontaneously broken.  
\end{enumerate}
We will reproduce this procedure in the following both for global and for local $\N=2$ supersymmetry.

\subsection{$\cN=2$ vector multiplet and new FI terms}

In this part we will describe the properties of the $\cN=2$ vector multiplet and introduce the new $\N=2$ Fayet--Iliopoulos term. 
To illustrate the properties of the construction we will break momentarily the manifest $SU(2)_R$ formulation and 
refer to the anticommuting coordinates as $\theta=\theta^1$ and 
$\tilde \theta = \theta^2$.\footnote{Here we use the conventions of \cite{Wess:1992cp}.} 
Later we will restore the manifest $SU(2)_R$ formulation but this first analysis might be useful to readers more familiar to $\cN=1$
superspace.

The abelian vector multiplet in global $\cN=2$ supersymmetry is described by a chiral superfield 
\be
\overline D_{\dot \alpha} W = 0 =  \overline {\tilde D}_{\dot \alpha} W \, , 
\ee 
that has a chiral $\tilde \theta$
 expansion\footnote{We set the magnetic FI parameter to zero as we are interested in a complete breaking of supersymmetry.}  
\be
\label{VM}
W = \Phi + \ri \, \tilde \theta^\alpha {\rm W}_\alpha(V) +  \tilde \theta^2  \left(- \frac14 \overline D^2 \overline \Phi   \right) \, . 
\ee
In \eqref{VM}  $\Phi$ is an $\cN=1$ chiral superfield and $V$ is an $\cN=1$ abelian gauge multiplet 
with 
\be 
{\rm W}_\alpha(V) = -\frac14 \overline D^2 D_\alpha V  \, , 
\ee  
its $\cN=1$ chiral spinorial field strength (where we use the abbreviations $D^2 = D^\alpha D_\alpha$, 
$\overline D^2 = \overline D_{\dot \alpha} \overline D^{\dot \alpha}$, 
$D \tilde D = D^\alpha \tilde D_\alpha$, etc.). 
The component fields of the $\cN=1$ chiral multiplet are defined as 
\be
\label{compPhi}
\Phi = A +  \theta \chi + \theta^2 {\rm F} \, , 
\ee
where $A$ is a complex scalar, 
$\chi$ is a Weyl spinor and ${\rm F}$ is a complex scalar auxiliary field. 
For the chiral field-strength superfield of the vector multiplet we have 
\be
\label{compW}
{\rm W}_\alpha(V) = - \ri \lambda_\alpha 
+ \Big{[} \delta_\alpha^\beta \text{D} - \frac{\ri}{2} (\sigma^m \overline \sigma^n)_\alpha^{\ \beta} F_{mn} \Big{]} \theta_\beta 
+ \theta^2 (\sigma^m)_{\alpha \dot \alpha} \partial_m \overline \lambda^{\dot \alpha} \, , 
\ee
where $F_{mn} = \partial_m v_n - \partial_n v_m$ for the abelian gauge field $v_m$, 
$\lambda$ is a Majorana spinor and D a real scalar auxiliary field. 
Eqs.~\eqref{VM}--\eqref{compW} are written in chiral coordinates. 
The action of ${\cal N}=2$ supersymmetry on $V$ implies that $\Phi = \overline{DD}\widetilde V$, where $\widetilde V$ 
is an ${\cal N}=2$ partner of $V$.\footnote{Note that 
in a projective superspace approach \cite{KLR,LR3,LR2} to off-shell $\cN=2$ supersymmetry,
 see \cite{Lindstrom:2008gs,Kuzenko:2010bd} for reviews,
the unconstrained prepotential for an $\cN=2$ Abelian vector multiplet  \cite{LR2}  is described
by an infinite series of $\cN=2$ superfields
$V_k(z)$ organized as Laurent series $\cV(z,\z)=\sum_{k=-\infty}^{+\infty}\z^kV_k(z)$ 
in terms of an auxiliary complex coordinate $\z$ such that 
$(\z D_\a^1- D_\a^2)\cV(z,\z)=(\DB^\ad_1+\z\DB^\ad_2)\cV(z,\z)=0$ and $V_k=(-)^k\overline{V}_{-k}$.
The field strength of the $\cN=2$ vector multiplet then satisfies 
$W=-\frac{\ri}{4}\DB_{\ad 1}\DB^\ad_1V_1=-\frac{\ri}{4}\DB_{\ad 2}\DB^\ad_2V_{-1}$ which, once reduced to $\cN=1$ superspace,
gives 
$\F=W|_{\q^2}=-\frac{\ri}{4}\DB^2V_1|_{\q^2=0}$ and 
${\rm W}_\a=-\ri D_\a^2W|_{\q^2=0}=-\frac{1}{4}\DB^2D_\a V_0|_{\q^2=0}$ that identifies
$\widetilde V=-\frac{\ri}{4}V_1|_{\q^2=0}$ and the $\cN=1$ vector multiplet prepotential as $V=V_0|_{\q^2=0}$.} 
This fact does not have consequences for our analysis here.

The two-derivative model for an $\cN=2$ vector multiplet is given by the superspace Lagrangian 
\be
\label{FSP}
\begin{aligned}
{\cal L}_{\text{kinetic}} =&\, \frac12 \int \rd^2 \theta\, \rd^2 \tilde \theta \, {\cal F}(W)  + c.c. 
\\[0.2cm]
=& \int \rd^2 \theta \,\rd^2 \bar \theta \left[ \frac12 {\cal F}'(\Phi) \overline \Phi + c.c. \right] 
+ \frac18 \left\{ \int \rd^2 \theta \, {\cal F}''(\Phi) \, {\rm W}^\alpha(V) {\rm W}_\alpha(V) + c.c. \right\} \, . 
\end{aligned}
\ee
The bosonic sector of \eqref{FSP} is 
\be
\label{KINB}
\begin{aligned}
{\cal L}_{\text{kinetic}}^{(bosons)} =& \  \text{Re}{\cal F}'' F \overline F 
- \text{Re}{\cal F}'' \partial_m A \partial^m \overline A \\ 
& + \frac14 \text{Re}{\cal F}'' \text{D}^2 
- \frac18 \text{Re}{\cal F}'' F_{mn} F^{mn} 
+ \frac{1}{16} \text{Im} {\cal F}'' F_{mn} F_{kl} \epsilon^{mnkl} \, ,  
\end{aligned}
\ee 
where 
\be
\text{Re} {\cal F}'' = \frac12 \left[ {\cal F}''(A) + \overline{\cal F}''(\overline A) \right] \, , \quad  
\text{Im} {\cal F}'' = \frac{1}{2\ri} \left[ {\cal F}''(A) - \overline{\cal F}''(\overline A) \right] \, , 
\ee 
and ${\cal F}'(A) = \frac{\partial {\cal F}(A)}{\partial A}$. 
We are interested in the study of spontaneous supersymmetry breaking 
and therefore only the shifts in the supersymmetry transformations of the 
fermions are relevant here. 
The fermion shifts have the form 
\be
\label{fermsusy}
\delta \lambda_\alpha  = \ri \text{D} \epsilon_\alpha 
+ 2  \overline {\rm F}  \eta_\alpha + \dots \, , \quad 
\delta \chi_\alpha  =  2 {\rm F} \epsilon_\alpha 
+ \ri \eta_\alpha \text{D} +\dots \, . 
\ee
From \eqref{fermsusy} we see that in general $\cN=2$ supersymmetry is spontaneously broken to $\cN=0$ if either 
auxiliary fields ${\rm F}$ or D acquire a {\it vev}. 
Therefore when supersymmetry is broken by the vector multiplet  it holds
\be
\label{vevs}
\langle {\rm F} \rangle \ne 0 \, ,  \quad \text{and/or} \quad \langle \text{D} \rangle \ne 0 \, . 
\ee 
The simplest way to achieve a setup where this is realized is by adding to \eqref{FSP}  Fayet--Ilioupoulos
terms of the form  \cite{Fayet:1974jb,Fayet:1975yi,VanProeyen:2004xt}
\be
\label{SFI}
{\cal L}_{\text{standard FI}} = - \xi \text{D} - f {\rm F} - \bar{f} \overline {\rm F} \, , 
\ee
for a real constant $\xi$ and a complex constant $f$. 
Once we integrate out the auxiliary fields we have 
\be
\langle \text{D} \rangle = \frac{2 \xi}{\text{Re}{\cal F}'' } \, , \quad \langle {\rm F} \rangle = \frac{\bar{f} }{\text{Re}{\cal F}'' } \, .  
\label{vacua}
\ee
Notice that the term \eqref{SFI} is invariant under supersymmetry, 
therefore it can be consistently added to the Lagrangian \eqref{FSP} and thus breaks supersymmetry only spontaneously. 
The scalar potential of the resulting theory is 
\be
\label{VAPT}
{\cal V} = \frac{|f|^2 + \xi^2 }{\text{Re} {\cal F}''} \, .  
\ee
Notice that within this set-up the previous scalar potential generically leads to a run-away behavior that will restore supersymmetry.
The only consistent setup is to have $\text{Re} {\cal F}''=$const. that leads to a constant scalar potential, 
though the Lagrangian will describe a non-interacting theory. 
Consistent interacting supersymmetry breaking patterns are known to arise if both Electric and Magnetic FI terms are turned on 
\cite{Antoniadis:1995vb} 
or when these models are coupled to supergravity. We will keep for simplicity the Magnetic FI terms turned off in this notes and focus 
on supergravity in the next sections.

Assuming that \eqref{vevs} holds we can derive a property for a specific composite superfield 
that will be relevant for the rest of our discussion. 
We have 
\be
\label{vev8}
\Big{\langle} 
 \left( D^4 \overline D^4 |DW|^8  \right) \Big{|}_{\theta=\tilde \theta =0} 
\Big{\rangle} 
= \Big{|} \Big{\langle} \left( 16 {\rm F} \overline {\rm F} + 4 \text{D}^2 \right)^2 \Big{\rangle} \Big{|}^2  \ne 0 \, , 
\ee
where 
\be
D^4 = D^2 \tilde D^2 \, , \quad 
\overline D^4 = \overline D^2 \overline{\tilde D}^2 
\, , \quad 
|DW|^8 = 
|D^\alpha W D_\alpha W|^2 
|\tilde D^\beta W \tilde D_\beta W|^2 \, . 
\ee
It is important to stress that the previous condition is equivalent to the requirement that the vacuum breaks completely
supersymmetry which, according to eqs.~\eqref{vevs} and \eqref{vacua}, is consistent whenever 
$\xi$ and/or $f$ are nonvanishing.
From \eqref{vev8} we then see that the superfield 
\be 
\Big{[} D^4 \overline D^4 
\left( |D W|^8 \right) \Big{]}^{-1} 
\, , 
\ee
is always well-defined as long as \eqref{vevs} holds, 
i.e., 
as long as the vector multiplet contributes to the {\it complete} supersymmetry breaking. 
We can now introduce the {\it new} $\cN=2$ Fayet--Iliopoulos term, 
which is given by the expression 
\be
\label{newSFI}
{\cal L}_{\text{new FI}} = \int \rd^8 \theta \, 
\frac{16^2 \,|DW|^8}
{D^4 \overline D^4  |DW|^8}  
\left\{ -\frac{\ri}{2} \xi D \tilde D W + \frac14 f D^2 W + \frac14 \bar{f} \overline D^2 \overline W \right\} \, . 
\ee
Note that the pre-factor $(16^2 \,|DW|^8)/(D^4 \overline D^4  |DW|^8)$ 
  in \eqref{newSFI} is chosen to pick the lowest component of the Lagrangian in the bracket
as the only bosonic part of the component action, similarly in spirit to \eqref{genU}. 
The bosonic sector of \eqref{newSFI} can be seen to match with \eqref{SFI}, 
therefore supersymmetry will be broken by the vector multiplet, 
thus making the term \eqref{newSFI} self-consistent.

As we explained earlier the scalar potential \eqref{VAPT} however leads to a runaway 
behavior that will restore supersymmetry unless the 
function $\text{Re} {\cal F}''$ is constant. 
In the setup with a standard FI-term this leads to a non-interacting theory. 
However, with the new FI term \eqref{newSFI}  and with constant $\text{Re} {\cal F}''$, 
the theory is interacting due to the higher order fermionic interactions of the type appearing in \eqref{genU}. 
In particular, the theory will contain the standard kinetic terms given in \eqref{KINB}, 
the related kinetic terms for the gaugini, 
and a series of higher order non-linear interactions that will always contain fermions 
and will be generically suppressed by the supersymmetry breaking scale $\sqrt{\cal V}$.

We can now recast our results in the $SU(2)_R$ covariant formulation 
and underline the properties of the non-linear structure of \eqref{newSFI}. 
The $\cN=2$ chiral multiplet constraints can be written 
in a covariant $SU(2)_R$ description and take the form \cite{Grimm:1977xp}
\be
\overline D_{\dot \alpha i} W = 0 \, , 
\ee
and 
\be
\label{DEFS}
D^{ij} W = \overline D^{kl} \overline W =\epsilon^{ik} \epsilon^{jl} \, \overline D_{kl} \overline W \, . 
\ee 
We have used the abbreviations 
\be
D^{ij} = D^{\a i} D_\a^j \, , \quad \overline D_{ij} = \overline D_{\ad i} \overline D^\ad_{j} \, , 
\ee
and we follow conventions where $\epsilon^{12} =-\epsilon_{12} =1$.\footnote{For the covariant $SU(2)_R$ notations
we follow \cite{Butter:2012xg}.} 
The auxiliary fields can be recast into an $SU(2)_R$ covariant notation by defining the symmetric and real isotriplet,
$X^{ij}=X^{ji}$, $\overline{(X^{ij})}=X_{ij}$, as
\be
X^{ij}=  D^{ij} W | =  \begin{pmatrix}
    -4 {\rm F} & -2 \ri \, \text{D}  \\
  -2 \ri \, \text{D}   &-4  \overline {\rm F}
  \end{pmatrix} \, 
  , \quad X_{ij}=  \begin{pmatrix}
   -4  \overline {\rm F} & ~2  \ri \, \text{D} \\
   2 \ri \, \text{D} &  -4 {\rm F}  
  \end{pmatrix} \,  , 
  \label{Xflat}
\ee
which gives $\frac{1}{16} X^{ij}X_{ij}=\frac{1}{16}\det[X^{ij}]= 2 |{\rm F}|^2 + \frac12 \text{D}^2$.
Note that $X^{ik} X_{kj} = \delta^i_j \, X^{pq} X_{pq} /2$ 
and that the fermions 
$\l_\a^i=(\chi_\a,\lambda_\alpha)$
shift under supersymmetry as $\delta \lambda^k_\alpha = -\frac12 \e_{j \alpha} X^{k j} 
+\dots$ 
Once we define 
\be
\Delta = \frac{1}{48} D^{ij} D_{ij} = \frac{1}{16} D^2 \tilde D^2 
\, , \quad 
\overline \Delta = \frac{1}{48} \overline D^{ij} \overline{D}_{ij}=\frac{1}{16} \overline{D}^2 \overline{\tilde D}^2  
\, , 
\label{chiral-projector-flat}
\ee 
we find\footnote{Note that in the $SU(2)_R$ covariant notation we have 
$|D W|^8  = \frac19 \Big{|} (D^{\a i} W) (D_\a^j W) (D^\b_i W) (D_{\b j} W)\Big{|}^2 $.} 
\be
\Big{\langle} 
\Delta \overline \Delta |D W|^8 
\Big{|}_{\theta_i =0} 
\Big{\rangle} 
= \Big{\langle} 
\left( \frac18 X^{ij}X_{ij}\right)^4 
\Big{\rangle} 
\, . 
\ee 
The complete breaking of supersymmetry is then equivalent to\footnote{ 
Magnetic Fayet--Iliopoulos terms can be described as deformations of the constraint \eqref{DEFS} by mean of a constant real isotriplet 
 $M^{ij}=\overline{(M_{ij})}$
 as $(D^{ij}W-\DB^{ij}\overline{W})=4\ri M^{ij}$; see 
 \cite{Antoniadis:1995vb,IZ1,IZ2,RT,Kuzenko:2015rfx,Antoniadis:2017jsk,Antoniadis:2019gbd,Cribiori:2018jjh} 
 for the case of $\cN=2$ global supersymmetry and \cite{Kuzenko:2013gva,Kuzenko:2015rfx,Kuzenko:2017gsc} 
 for extensions to curved $\cN=2$ superspaces and local supersymmetry.
 In this case $X^{ij}=D^{ij}W|_{\q=0}$ is not real any longer and it is possible to have cases where 
at last one of the components of $\langle X^{ij} \rangle$ is non-zero but $\langle X^{ij}X_{ij}\rangle\equiv 0$. 
In this case supersymmetry is spontaneously broken from $\cN=2$ to $\cN=1$.}  
\be
\langle X^{ij}X_{ij}\rangle  \ne 0 \, . 
\label{XXnon0}
\ee
We can now recast \eqref{newSFI} in the form 
\be
\label{SU2FI}
{\cal L}_{\text{new FI}} = \int \rd^8 \theta \, 
\frac{   |D W|^8  }
{\Delta \overline \Delta  |D W|^8   }  
\left\{ \frac18 \xi_{ij} D^{ij} W + c.c. \right\} \, , 
\ee
which delivers in the bosonic sector 
\be
\label{compSU2}
{\cal L}^{(bos.)}_{\text{new FI}} = \frac18 \xi_{ij} X^{ij}+ c.c.  
\ee
To match \eqref{SU2FI} to \eqref{newSFI} (or equivalently \eqref{compSU2} to \eqref{SFI}) we can set 
\be
\xi_{ij} =  \begin{pmatrix}
   f  & -\ri\, \xi  \\
  -\ri\, \xi  &  \bar{f}    
  \end{pmatrix} \, , 
\ee 
which is however not a unique choice as there is the freedom of $SU(2)_R$ rotations.

Now we are ready to relate the Lagrangian \eqref{SU2FI} to the underlying goldstino structure. 
We start by defining the following composite nilpotent chiral superfield
\be
\label{Xsusy}
X= \overline \Delta \left( |D W|^8   \right) \, , 
\ee
which has the properties 
\be
 X^2 = 0 \, , \quad \overline D_{\ad}^i {X} = 0 \, , \quad \langle \Delta {X} \rangle \ne 0 
~, 
\label{nilpotencyX1}
\ee
with the last one holding only when supersymmetry is completely broken, 
i.e. when \eqref{XXnon0} holds. 
By using the results  in the appendix A, 
besides $X^2=0$, 
the $X$ superfield can be shown to satisfy by construction a series of nilpotency conditions of the form 
\cite{Cribiori:2016hdz,Kuzenko:2017zla} 
\be
X D_AD_B X= 0~,~~~X D_AD_BD_C X= 0
 \, , \quad  D_A= (\pa_a,D_\alpha^i,D_i^{\dot\alpha}) \, . 
\label{nilpotencyX2}
\ee
As a result, 
one can also show that 
\be
\label{XcalX}
X \overline{X} = \overline \Delta \left( |D W|^8  \right)  
\Delta \left( |D W|^8  \right)  = \left( |D W|^8  \right) 
\overline \Delta  \Delta 
\left( |D W|^8 \right)  = \left( |D W|^8 \right) \Delta {X} \, . 
\ee
As a rapid cross-check, the reader can act on the two sides of \eqref{XcalX} with $\Delta$ and check that it gives an identity. 
From \eqref{XcalX} we derive 
\be
\label{ident}
\frac{  |D W|^8 }{ \Delta 
\overline \Delta |D W|^8} 
= \frac{{X} \overline{X} }{ \Delta {X} \overline \Delta \overline{X}} \, . 
\ee
Notice that the left-hand-side of \eqref{ident} is identical to the factor appearing in \eqref{SU2FI}. 
We can simplify \eqref{SU2FI} even further by relating to the $\Gamma_i$ goldstino superfields. 
Following \cite{Cribiori:2016hdz,Kuzenko:2017zla},  we know that from a nilpotent chiral superfield satisfying 
\eqref{nilpotencyX1} and \eqref{nilpotencyX2} it is always possible to define
the goldstini  superfields $\G_{\a i}$ of section \ref{section-goldstini-flat} as composite of $X$ as follows
\be
\label{Gsusy}
\Gamma_{\alpha i} = - \frac{1}{12} \frac{D^{j}_\alpha D_{ij} {X} }{\Delta {X}} \, .
\ee
By using the composite nilpotent chiral superfield defined in \eqref{Xsusy} we can then define
a Goldstino multiplet as a composite of the vector multiplet $W$, $\G_{\a i}=\G_{\a i}(W)$.
Its component field proves to be completely determined in terms of the vector multiplet components
\be
\gamma_{\alpha i} = -4 \frac{X_{ij}}{X^{pq} X_{pq}} \, \lambda^j_\alpha  + \dots \, ,
\label{flat-composite-goldstino}
\ee
where we have neglected in the previous equation terms that are function of $F_{mn}$ and derivatives of the
vector multiplet fields, 
or higher order than linear in the gaugini. 
The composite $\Gamma_{\a i}$ goldstino superfields have the property 
\be
\G^4 = \frac{X}{ \Delta X} \, , 
\quad 
\overline{\G}^4 
= \frac{\overline X}{ \overline\Delta \overline X} \, , 
\ee
which can be proven by using the nilpotency properties of $X$ given in \eqref{nilpotencyX2}. 
The above results mean that the new $\cN=2$ Fayet--Iliopoulos term can be recast in the equivalent form 
\be
\label{FIgamma}
{\cal L}_{\text{new FI}} =  \,  \int \rd^8 \theta \, 
|\Gamma|^8
\left\{ \frac18 \xi_{ij} D^{ij} W + c.c. \right\} \, , 
\ee
where the $\Gamma$ superfields are the {\it composite} objects that are uniquely defined in terms of $W$ from the 
procedure we presented above. 
The form of the Lagrangian \eqref{FIgamma} is exactly of the form \eqref{genU} that we analyzed earlier. 
As a result when we expand \eqref{FIgamma} in components we will find a bosonic sector given by $\frac18 \xi_{ij} D^{ij} W| + c.c.$  
and the rest will be terms proportional to the goldstini. 
More importantly, 
the form of the Lagrangian \eqref{FIgamma} is such that its embedding in $\cN=2$ supergravity can be achieved 
by following the results of \cite{Kuzenko:2017zla}.

Let us close this section with an observation on other possible deformations of the theory. 
Clearly because of the explicit introduction of non-linear realizations the deformations are numerous. 
First of all, it is clear that we could introduce in \eqref{FIgamma} an arbitrary function $H(W,\overline{W})$ 
of $W$ and $\overline{W}$ obtaining other types of FI terms
\be
\label{moreFIgamma}
{\cal L}_{\text{other FIs}} =  \,  \int \rd^8 \theta \, 
|\Gamma|^8
\,H(W,\overline{W})\,\xi_{ij} D^{ij} W + c.c. 
\ee
which, for simplicity, we will not investigate further both in the globally and locally supersymmetric cases.
Another simple example is to have a term of the form 
\be
\label{upliftglobal}
{\cal L}_{\text{Uplift}} = -  \int \rd^8 \theta \, 
|\Gamma|^8 \, U(W, \overline W) \, , 
\ee 
where the function $U(W,\overline{W})$ of $W$ and $\overline{W}$ is in general completely unconstrained.
This {\it uplift} term can of course only be sustained once supersymmetry is broken by the vector multiplet 
$W$ because it is the non-vanishing vevs of the auxiliary $X_{ij}^{(W)}$ 
that guarantee the self-consistency of the 
construction of the composite $\Gamma$ superfields. 
This means that in general an uplift term as \eqref{upliftglobal} has to come together with a term, such as the new FI term, 
that guarantees $\langle X^{ij}{}^{(W)}X_{ij}^{(W)}\rangle\ne0$.
From \eqref{genU} we see that the term \eqref{upliftglobal} leads to a contribution $U(A,\overline A)$ 
to the potential of the vector multiplet scalar fields, $A$ and $\overline A$, of the $\cN=2$ effective theory. 
Notice finally that a different type of deformations, 
that do not rely on non-linear realizations is possible. 
We could have also considered a term of the form 
\be
\int \rd^8 \theta \frac{W^2 \overline W^2}{\Delta W^2 \overline \Delta \overline W^2 } \, \xi_{ij} D^{ij} W \, , 
\ee
which would generate linear terms in $X^{ij}$. 
Such term however would also generate all sorts of higher derivative terms, 
for example terms including $\Box W^2 \Box \overline W^2$, 
that would not only lead to a complicated expression for the bosonic sector, 
but would possibly lead to ghost excitations within the effective theory. 
For this reason we neglect this kind of terms that at first sight might look as a natural $\cN=2$ generalization
of the $\cN=1$ new FI term of \cite{Cribiori:2017laj}.

\section{$\N=2$ Supergravity coupled to abelian vector multiplets}  
\label{review-off-shell-sugra}

We will now review some results about $\cN=2$ gauged supergravity constructed by using an off-shell setting
that might not be familiar to all the readers.
By following \cite{Butter4DN=2,Butter:2012xg}, we will also introduce superspace results that we will use for the rest of this work. 
We are going to employ an off-shell superconformal approach; 
see \cite{FVP} for a comprehensive review and also \cite{Butter4DN=2,Butter:2012xg} 
for $\cN=2$ conformal superspace, where Poincar\'e supergravity is obtained by coupling the Weyl multiplet of conformal supergravity 
to two compensators. 
We choose to use an off-shell setting where the two compensators are respectively
an $\N=2$ vector multiplet and an $\N=2$ tensor multiplet. 
For simplicity in this paper we focus on studying supergravity-matter couplings comprising only physical 
 vector multiplets without any physical hypermultiplets. In this section we start by introducing the superconformal multiplets
that will play a role in our discussion, 
then we describe the action associated to a generic system of vector multiplets coupled to off-shell $\cN=2$
Poincar\'e supergravity. We first consider the case of ungauged supergravity. After that we will 
explain how the standard FI term leads to gauged supergravity starting from an off-shell setting. 
Then we will start describing some properties of the vacuum structure of gauged supergravity focusing, 
in particular, to a model based on a single physical vector multiplet.
Though this section does not contain any original results it sets the stage to understanding the physical implications
 in $\cN=2$ supergravity that the new FI terms have compared to the standard one.

\subsection{The off-shell superconformal multiplets}

The Weyl multiplet  of $\cN=2$ conformal supergravity
is associated with the local off-shell gauging of the
superconformal group $SU(2,2|2)$ \cite{sct_rules,BdeRdeW,sct_structure}.\footnote{See also \cite{Cribiori:2018xdy}
for a recent description of the $\cN=2$ Weyl multiplet by using the rheonomic approach.}
The multiplet contains $24+24$ off-shell physical components described by a set of independent gauge fields:
the vielbein $e_m{}^a$ and a dilatation connection $b_m$;
the gravitino $\psi_m{}_i$, associated with the gauging of $Q$-supersymmetry;
and $U(1)_R\times SU(2)_R$ gauge fields $A_m$ and $\phi_m{}^{ij}$, respectively.
The other gauge fields associated with the Lorentz ($\o_m{}^{bc}$), 
special conformal ($\frak f_m{}^a$), 
and $S$-supersymmetry ($\phi_m{}^i$)
transformations of $SU(2,2|2)$ are composite fields.
In addition to the independent gauge connections, the Weyl multiplet comprises
a set of  covariant matter fields: a real rank two antisymmetric tensor $W_{ab}$; a real scalar field $D$;
and  a  fermion $\S^i$. These fields are necessary to close the algebra of local superconformal transformations 
without imposing equations of motion.

The field content of the $\cN=2$ Weyl multiplet 
can be embedded in a conformal superspace geometry 
as described in \cite{Butter4DN=2,Butter:2012xg} (we will closely follow the notation 
of \cite{Butter:2012xg}\footnote{The definition of the $(\sigma^{ab})_{\alpha}{}^\beta$ matrices in \cite{Ideas,Butter:2012xg} has 
an overal minus 
sign difference with the definition in \cite{Wess:1992cp}.}); 
see also appendix \ref{AppendixConfSuperspace} for a review of the results needed 
in our discussion.
The gauge fields of the $\cN=2$ Weyl multiplet are provided by the lowest components of the appropriate super one-forms 
\cite{Butter:2012xg}.
The vielbein ($e_m{}^a$) and gravitini ($\psi_m{}_i$) appear as the $\q=0$ projections of the coefficients of
$\rd x^m$ in the supervielbein $E^A$ one-form, 
that is 
\begin{align}
e{}^a = \rd x^m e_m{}^a = E^a\doubar~,~~~~~~
\psi^\a_i = \rd x^m \psi_m{}^\a_i =  2 E^\a_i \doubar ~,~~~
\bar\psi_\ad^i = \rd x^m \bar\psi_m{}_\ad^i =  2 \overline{E}_\ad^i \doubar ~,
\end{align}
where the double bar denotes setting $\q = \rd \q = 0$ \cite{Butter:2012xg,Baulieu:1986dp,BGG01}.
The remaining fundamental and composite one-forms correspond to
projections of superspace one-forms,
\bea
	A := \Phi \doubar~,~~
		\phi^{kl} := \Phi{}^{kl} \doubar~,~~
	b := B\doubar ~,~~
 \omega^{cd} :=  \Omega{}^{cd} \doubar ~,~~
{ \phi}_\g^k := 2 \,{ {\frak F}}{}_\g^k\doubar~, ~~
{\bar \phi}^\gd_k := 2 \,{\overline {\frak F}}{}^\gd_k\doubar~, ~~
 {\frak{f}}{}_c :=  {\frak{F}}{}_c\doubar~.
 ~~~
\eea
For instance, the spin connection $\o_m{}^{bc}$ is as usual composite and satisfies
\bea
 {\o}_{a}{}_{bc}
&=&
\omega(e)_{a}{}_{bc}
-2\eta_{a[b}b_{c]}
+{\rm fermions}
~,
\eea
where $\omega(e)_{a}{}_{bc} = \frac{1}{2} (\cC_{abc} + \cC_{cab} - \cC_{bca})$
is the torsion-free spin connection given
in terms of the anholonomy coefficient
$\cC_{mn}{}^a := 2 \,\partial_{[m} e_{n]}{}^a$.
In the following we will also use the expression for the trace of the special conformal connection $\frak f_m{}^a$,
which is also a composite field such that its trace satisfies
\be
{\frak f} = e_a^{\ m} {\frak f}_m^{\ a} = - D - \frac{1}{12} R(e,\omega) + {\text{fermions}} \, .
\ee
Here $R=R_{ab}{}^{ab}$ is the Ricci scalar with the Riemann tensor $R_{ab}{}^{cd}$ given by
\bea
R_{ab}{}^{cd}(\o)
=
e_a{}^me_b{}^n\Big(
2\pa_{[m}\o_{n]}{}^{cd}
+2\o_{[m}{}^{ce} \o_{n]}{}_e{}^d
\Big)
~.
\eea

The covariant auxiliary fields of the Weyl multiplet, 
$W_{ab}$, $D$, and $\S^i$,
 belong to some of the components of the primary $\N=2$ Weyl superfield $W_{ab}$
and its descendants.\footnote{A superfield $U$ is said to be \emph{primary} of dimension $\D$ if 
$K_a U =S^\a_i U=\overline{S}_\ad^i U= 0$,
and $\mathbb D U = \D U$.
The super-Weyl tensor $W_{ab}$ is a primary dimension 1 covariant superfield.}
In particular, the $\theta=0$ component of $W_{ab}$, $W_{ab}|_{\q=0}$,\footnote{We will often drop the $|_{\q=0}$ projection as it will
 be clear from the context when we consider a superfield, such as $W_{ab}$, or its lowest component, as $W_{ab}|_{\q=0}$}
describes the real rank-two matter field of the Weyl multiplet 
which is decomposed in its imaginary-(anti)-self-dual components,
$\frac{\ri}{2} \eps_{ab}{}^{cd} W_{cd}^{\pm} = \pm W_{ab}^{\pm} $,
 as
\bsubeq
\bea
&W_{ab} = W_{ab}^{+} + W_{ab}^{-} \ , \quad W_{ab}^{+} := (\s_{ab})^{\a\b} W_{\a\b} 
\ , \quad 
W_{ab}^{-} := - (\overline{\s}_{ab})_{\ad\bd} \overline{W}^{\ad\bd} \ ,\\
&W_{\alpha \beta} = \frac12 (\sigma^{ab})_{\alpha \beta} W_{ab} 
\, , \quad 
\overline W_{\dot \alpha \dot \beta} = -\frac12 (\overline{\sigma}^{ab})_{\dot \alpha \dot \beta} 
W_{ab} 
= \overline{(W_{\alpha \beta})} \, .
\eea
\esubeq
The self-dual and anti-self-dual parts of $W_{ab}$ have different transformations under the chiral $U(1)_R$ symmetry: 
$YW_{\a\b}=-2W_{\a\b}$ and $Y\overline{W}_{\ad\bd}=2\overline{W}_{\ad\bd}$.
The fermion and the other real covariant field of the Weyl multiplet ($\Sigma^i$ and $D$) 
are associated with the projections 
\be
\Sigma^{\alpha i} 
= 
\frac13 \nabla^i_\beta W^{\alpha \beta}| \, , \quad 
\overline{\Sigma}_{\ad i} 
= 
-\frac13 \overline{\nabla}_i^\bd \overline{W}_{\ad \bd}| \, , \quad 
D = \frac{1}{12} \nabla^{\alpha \beta} W_{\alpha \beta}|
= \frac{1}{12} \overline\nabla^{\ad \bd} \overline{W}_{\ad \bd}| \, ,  
\ee 
where
\be
\nabla^{ij} = \nabla^{\gamma (i} \nabla_{\gamma}^{j)} \, , \quad \overline \nabla^{ij} 
= \overline \nabla_{\dot \gamma}^{(i} \overline \nabla^{j) \dot \gamma} 
\, ,
\quad
\nabla_{\a\b} = \nabla^{k}_{(\a} \nabla_{\b) k}
 \, , \quad 
 \overline \nabla_{\ad\bd} 
= \overline \nabla_{(\dot \a k} \overline \nabla_{\dot \b)}^k \, . 
\ee
The algebra satisfied by the $\N=2$ conformal superspace derivatives $\de_A=(\de_a,\nabla_\alpha^i,\deb^\ad_i)$ 
can be found in appendix B. 
For the reader familiar with the superconformal techniques \cite{FVP} it might be useful to underline that in the conformal 
superspace framework the spinor derivatives $\nabla_\alpha^i$ and $\deb^\ad_i$ play 
the role of the $Q$-supersymmetry generators $Q_\alpha^i$ and $\overline{Q}^\ad_i$ while the vector derivative 
$\de_a$ is, as usual, associated to the momentum operator $P_a$ of the soft algebra describing the gauging of the
superconformal algebra. 
More precisely, given a covariant tensor superfield $T$, 
it will transform under local $SU(2,2|2)$ transformations as\footnote{As also 
described in appendix \ref{AppendixConfSuperspace}, $J_{ij}$, $Y$ and $\mathbb{D}$ are the 
$SU(2)_{R}$, $U(1)_{R}$ and dilatation generators respectively
while $K^a$ is the special conformal generator, and $(S^\a_i,\,\overline{S}_\ad^i)$ are the $S$-supersymmetry generators
that for convenience are grouped together  as $K^A=(K^a,S^\a_i,\overline{S}_\ad^i)$. }
\be
\d_{\cG} T =
 \cK T
~,~~~
\cK = \xi^C  {\nabla}_C + \hf  {\L}^{ab} M_{ab} +  {\L}^{ij} J_{ij}+\ri\t Y +  \s \mathbb D +  {\L}_A K^A ~. 
\ee
The projected spinor covariant derivatives $\nabla_{\a}^i\loco$ and $\overline{\nabla}^{\ad}_i\loco$
correspond to the $Q$-supersymmetry generator, and are defined so that, for example,
given the component field $\cT = T\vert=T\vert_{\q=0}$, then the action of the $Q$-supercharge is defined as 
$Q_\a^i\cT:=\nabla_\a^i| \cT := (\nabla_\a^i T) \loco$, etc. 
For the other generators, the action on the component field $\cT$ 
is simply given by the projection of the superfield analogue as e.g.~$M_{c d}\cT =(M_{c d} T)\loco$.
By taking the component projection of the superform $\de=E^A\de_A$,
the component vector covariant derivative ${\de}_a$ is defined to coincide with
the projection of the superspace derivative\footnote{Similarly to $W_{ab}$, we will use ${\nabla}_a$ 
to denote both the superspace or the component vector derivatives since it will be clear from the context
which one we are referring to.}
 ${\nabla}_a \loco$
\begin{align}
e_m{}^a \nabla_a \lc &= \partial_m
	- \hf \psi_m{}^\g_k \nabla_\g^k\lc - \hf \bar{\psi}_m{}_\gd^k \overline{\nabla}^\gd_k\lc
	+ \hf \omega_m{}^{bc} M_{bc}+ \phi_m{}^{ij} J_{ij} + \ri A_m Y + b_m \mathbb{D}
	\eol & \quad
	+ \frak{f}_m{}^b K_b + \hf \phi_m{}^i_\a S^\a_i + \hf \bar\phi_m{}^i_\ad \overline{S}^\ad_i~.
\end{align}
The component supercovariant curvature tensors,
arising from the commutator of two $ \de_a$ derivatives,
then coincide with the lowest components of the corresponding superspace curvatures.
The component and conformal superspace formalisms then prove to be equivalent with the difference that 
in the latter case local supersymmetry is geometrically realized as the spinor component of superdiffeomorphisms.
In the following, when we discuss component fields, we will also use the derivative
\be
\nabla'_a = {\rm D}_a(\omega) + \phi_a^{\ ij} J_{ij} + \ri A_a Y + b_a \mathbb{D} \, , 
\ee
where 
\be
{\rm D}_a(\omega) = e_a^m \left( \partial_m + \hf\omega_m^{\ ab} M_{ab} \right) \, . 
\ee
When we gauge fix the special conformal transformations we choose $b_a=0$. 
We refer to \cite{Butter:2012xg,Butter4DN=2} for a detailed discussion about the relation between $\cN=2$ conformal superspace
and the standard superconformal tensor calculus techniques and, in particular, 
for the supersymmetry transformations of the components of the Weyl multiplet
which are not needed for the scope of this paper. 

Let us now turn to the description of the matter multiplets embedded in a conformal supergravity setup. 
For the matter and the compensator sector we will work with $\N=2$ vector and 
tensor multiplets. 
The definition of an abelian vector multiplet in our setup is 
\be
\label{DEFW}
\overline \nabla^i_{\dot \alpha} {W}  = 0 \, , \quad \nabla^{ij} {W} = \overline \nabla^{ij} \overline{W} \, , 
\ee
where ${W}$ is a chiral primary complex scalar superfield ($K^A{W}=0$) with weights 
\be
\mathbb{D} \, {W} = {W} \, , \quad Y \, {W} = - 2 \, {W} \, . 
\label{DWYW}
\ee
The component fields of the vector multiplet are the complex scalar $\phi$, 
the gaugini $\lambda^i_\alpha$  and the $SU(2)_{R}$ triplet of auxiliary fields  $X^{ij}$, 
which are defined as 
\be
\label{CompVect}
\phi = {W}| \, , \quad \lambda_\alpha^i = \nabla_\alpha^i {W} | \, , \quad X^{ij} = \nabla^{ij} {W} | \, , 
\ee
whereas the field strength of the abelian gauge field  resides in the component 
\be
- \frac18 (\sigma_{ab})_{\alpha \beta} (\nabla^{\alpha \beta} {W} + 4 {W}^{\alpha \beta}  \overline{W} ) \Big{|} 
+ \frac18 (\overline \sigma_{ab})_{\dot \alpha \dot \beta} (\overline \nabla^{\dot \alpha \dot \beta} \overline{W} 
+ 4 \overline{W}^{\dot \alpha \dot \beta}  {W} ) \Big{|}  
= F_{ab} + {\text{fermions}} \, , 
\ee
where $F_{ab} = e_a^m e_b^n (\partial_m v_n - \partial_n v_m)$.

The off-shell $\cN=2$ tensor \cite{N=2tensor,Siegel:1978yi,Siegel80,SSW,LR3} 
(or also called linear) multiplet 
coupled to conformal supergravity, which will only play the role of a compensator in our 
paper,  is described by a superfield \cite{Butter:2012xg} ${\cal G}^{ij}={\cal G}^{ji}$
which is a primary ($K^A {\cal G}^{ij}=0$)
with the following dilatation and $U(1)_R$ weights 
\be
\mathbb{D} \, {\cal G}^{ij} = 2 {\cal G}^{ij} \, , \quad Y \, {\cal G}^{ij} = 0 \, , 
\ee
and satisfies the conditions 
\be
\overline{({\cal G}^{ij})} = {\cal G}_{ij} \, , \quad 
\overline \nabla_{\dot \alpha}^{(i}  {\cal G}^{jk)} = 0 \, , \quad  
\nabla_{\alpha}^{(i}  {\cal G}^{jk)} = 0 \, . 
\ee
The tensor multiplet constraints can be solved as
\be
{\cal G}^{ij} = \frac14 \nabla^{ij} \Psi + \frac14 \overline \nabla^{ij} \overline \Psi \, , 
\label{prepotential-tensor}
\ee
where $\Psi$ is an unconstrained $\cN=2$ chiral primary superfield with weights
\be
\mathbb{D} \, \Psi = \Psi \, , \quad Y \, \Psi= - 2 \, \Psi \, . 
\ee
The covariant component fields that reside in the tensor multiplet are given by 
\be
G^{ij} = {\cal G}^{ij}| \, , \quad 
\chi_{\alpha i} = \frac13 \nabla_\alpha^j {\cal G}_{ij} | \, , \quad 
F = \frac{1}{12} \nabla^{ij} {\cal G}_{ij} | \, . 
\ee
The real scalars $G^{ij}$ form an  $SU(2)_{R}$ triplet whereas $F$ is a complex scalar singlet. 
We will use the abbreviations 
\be
{\cal G} = \sqrt{ {\cal G}^{ij} {\cal G}_{ij} /2 } \, , \quad G = \sqrt{ G^{ij} G_{ij} /2 } \, . 
\ee 
From the prepotential $\Psi$ we also obtain the gauge  two-form $b_{mn}$ of the tensor multiplet as follows \cite{Butter:2012xg} 
\be
b_{mn} e_a^{\ m} e_b^{\ n} = B_{ab}| = - \frac{\ri}{4} (\sigma_{ab})^{\alpha \beta} (\nabla_{\alpha \beta} \Psi
 - 4 {W}_{\alpha \beta}  \overline{\Psi} ) \Big{|} 
- \frac{\ri}{4} (\overline \sigma_{ab})_{\dot \alpha \dot \beta} (\overline \nabla^{\dot \alpha \dot \beta} \overline{\Psi} 
- 4 \overline{W}^{\dot \alpha \dot \beta}  {\Psi} ) \Big{|}  \, . 
\ee
The two-form will usually appear through its supercovariant field strength 
\be
\tilde h^a = \frac16 \ve^{abcd} h_{bcd} + {\rm fermions}  \, , \quad   h_{mnp} = 3 \partial_{[m} b_{np]} \, . 
\ee

\subsection{Ungauged $\cN=2$ supergravity}

We can now describe actions for two-derivative matter-coupled Poincar\'e supergravity within an off-shell setting. 
We first look at ungauged $\cN=2$ supergravity. 
We consider a system of Abelian vector multiplets coupled to 
$\N=2$ conformal supergravity 
\be
{W}^I = ( {W}^0 , {W}^A ) \, ,  
\ee
where ${W}^0$ will serve as compensator and ${W}^A$ are the physical ones. 
We consider the following Lagrangian 
\be
\label{standardSG}
{\cal L}_{\rm ungauged} = \left\{ - {\cal L}_{\rm V} + c.c. \right\} 
+ {\cal L}_{\rm L} 
\, . 
\ee

The ${\cal L}_{\rm V}$ part describes the coupling of conformal supergravity 
to the physical and compensator vector multiplets.
It arises from the following chiral superspace action
\be 
S_{\rm V} = S_c + {\rm c.c.} \ , \quad 
S_c = \int \rd^4 x \, \rd^4 \q \,\cE \, \cF({W}^I) \ ,
\label{SV}
\ee
where the special-K\"ahler prepotential $\cF({W}^I)$ 
is holomorphic and homogeneous of degree two
\be
\frac{\pa \cF}{\pa \overline{{W}}^I}=0
~,~~~~~~
{W}^I {\cF}_{I}={W}^I \frac{\pa \cF}{\pa {W}^I}=2\cF \, . 
\ee
This guarantees that $\cF({W}^I)$ is a chiral primary with conformal dimension two and $U(1)_R$ weight $-4$, 
and then the action $S_c$ is locally superconformal invariant \cite{Butter4DN=2}.
The bosonic sector of \eqref{SV} appearing in \eqref{standardSG} in our notation is given by \cite{Butter:2012xg}
\be
\begin{aligned}
e^{-1} \left\{ - {\cal L}_{\rm V} + c.c. \right\}  = & - {\cal F}_I \Box \overline \phi^I 
- \frac{1}{32} {\cal F}_{IJ} X^{Iij} X^J_{ij} 
+ 2 {\cal F}_{IJ} F^{I \alpha \beta} F_{\alpha \beta}^J 
+{\cal F} \overline W_{\dot \alpha \dot \beta} \overline W^{\dot \alpha \dot \beta} 
- 3 D {\cal F}_I \overline \phi^I 
\\
& + 2 {\cal F}_I \overline W_{\dot \alpha \dot \beta} F^{I \dot \alpha \dot \beta} 
+ 2 {\cal F}_{IJ} \overline \phi^I W^{\alpha \beta} F_{\alpha \beta}^J 
+ \frac12 {\cal F}_{IJ} \overline \phi^I \overline \phi^J W^{\alpha \beta} W_{\alpha \beta} 
+ c.c. 
\,,
\end{aligned}
\ee
where the superconformal d'Alembertian is
\be
\Box \overline \phi^I = {\nabla'}^a{\nabla'}_a \overline \phi^I + 2 {\frak f} \, \overline \phi^I \, .  
\ee

The ${\cal L}_{\rm L}$ Lagrangian describing  the action for the 
tensor multiplet compensator coupled to conformal supergravity 
can be obtained from the following conformal superspace chiral action
\begin{align}\label{PsiW1}
S_{\rm L} = \int \rd^4x\, \rd^4\q\, \cE\, \Psi \mathbb W  + \HC \, , 
\end{align}
where $\Psi$ is the prepotential for the tensor multiplet
and $\mathbb W=\mathbb W[\cG]$ 
is a composite vector multiplet field strength constructed in terms of the tensor multiplet $\cG^{ij}$.
Its form is given by \cite{Muller:ChiralActions,Butter:2012xg}
\be\label{compositeVector}
\mathbb W = - \frac{1}{24 \cG} \overline{\nabla}_{ij} \cG^{ij} 
+ \frac{1}{36 \cG^3}\cG_{ij} \overline{\nabla}_{\ad k} \cG^{ki} \overline{\nabla}^\ad_l \cG^{lj}  \ .
\ee
The action \eqref{PsiW1} is a conformal superspace version of the improved tensor multiplet action 
 \cite{deWPV, LR:ScalarTensor,Siegel85}.
Its bosonic sector is given by the  Lagrangian \cite{Butter:2012xg}
\be
\begin{aligned}
e^{-1}  {\cal L}_{\rm L} = 
& - \frac{1}{2 G} |F|^2 
+ \frac14 G_{ij} \frac{1}{G} \left(- 2 \Box G^{ij} - 6 G^{ij} D \right) 
- \frac12 \epsilon^{mnpq} b_{mn} f^{\rm L}_{pq}
\\
& + \frac14 G_{ij}  \frac{1}{G^3} 
\left( 
{\nabla'}^a G^{ik} {\nabla'}_a G^{jl} G_{kl} 
+ \tilde h^a \tilde h_a G^{ij} 
-2 \tilde h^a {\nabla'}_a G^{k(i} G^{j)}_k \right) 
 \, ,
\end{aligned}
\ee
where the bosonic parts of the superconformal d'Alembertian is given by
\be
\Box G^{ij} = {\nabla'}_a {\nabla'}^a G^{ij} + 4 {\frak f} G^{ij}  
~,
\ee
while the composite two-form $f^{\rm L}_{mn}$ is
\be
f^{\rm L}_{mn}  = \partial_{m} \Big{[} \frac{1}{2G} \phi_n{}^{ij} G_{ij} + \frac{1}{2G} e_n{}^{a} \tilde h_a \Big{]} 
- \partial_n \Big{[} \frac{1}{2G} \phi_m{}^{ij} G_{ij} + \frac{1}{2G} e_m{}^{a} \tilde h_a \Big{]} 
+ \frac{1}{4G^3} \partial_m G^{ik} \partial_n G_k^{\ j} G_{ij}  \, . 
\ee

The dynamical system described by \eqref{standardSG} includes several auxiliary fields 
and pure gauge degrees of freedom that can be eliminated algebraically 
to obtain on-shell $\N=2$ Poincar\'e supergravity. 
We focus our attention only on the bosonic fields and start our discussion from the scalar and gravitational sector. 
First we integrate out the auxiliary field $D$ which gives 
\be
\delta D \quad \Longrightarrow \quad N = G \, , 
\label{varD}
\ee
where $N$ defines the special-K\"ahler potential
\be
N = {\cal F}_I \, \overline \phi^I + \overline{\cal F}_I \, \phi^I \, , \quad N_{IJ} = {\cal F}_{IJ} + \overline{\cal F}_{IJ} \, . 
\ee 
The tensor $N_{IJ}$ is generically chosen to have $(d,1)$ Lorentzian signature, 
where $d$ is the number of physical vector multiplets 
and the single positive signature direction indicates the presence of a compensator, 
here chosen to be $\phi^0$, among the vector multiplets.
By imposing that the physical fields have canonical
kinetic terms, the signature requirements on $N_{IJ}$ have been discussed for example in \cite{Cremmer:1984hj}, 
and we will see how it is respected by our examples. 
If we demand the Einstein--Hilbert term to be canonically normalized, $\frac12 eR$,
then we also need to impose the dilatation gauge fixing condition 
\be
\label{GaugeFix}
N=G={\cal F}_I \, \overline \phi^I + \overline{\cal F}_I \, \phi^I =1 \, . 
\ee
Part of the $SU(2)_R$ symmetry is fixed for convenience by imposing the gauge condition
\bea
G^{ij} = \delta^{ij} 
\, ,
\label{GaugeFix-2}
\eea
which leaves an off-shell residual $\widehat{U}(1)_R$ symmetry gauged by the following connection
\bea
\hat \phi_m:=\phi_m{}^{ij}\d_{ij}
~.
\eea
Next, we integrate out the auxiliary field $A_m$ which gives 
\be
\delta A_m  \quad \Longrightarrow \quad A_m = \frac{\ri}{4} N_{IJ} X^I \partial_m \overline X^{J} 
- \frac{\ri}{4} N_{IJ} \overline X^{I} \partial_m X^{J} 
~,
\label{varA}
\ee
and we fix the compensating scalar $\phi^0$  as a function of the other scalar fields
by imposing the condition \eqref{GaugeFix}. 
The previous result describes how the gauging of the K\"ahler transformations is identified with the $U(1)_R$ symmetry.
At this point, we can recast the kinetic terms of the scalar and gravitational sector in 
the standard form 
\be
\label{VST}
e^{-1} {\cal L}^{\text{ungauged}}_{\text{scalar-gravity}} 
=
 \frac12 R - \sum_{I,J \ne 0} g_{I \overline J} \, \partial_m z^I \partial^m \overline z^J 
\, , 
\ee
where $g_{I \overline J}$ is the K\"ahler metric 
\be
g_{I \overline J} = \frac{\partial^2}{\partial z^I \partial \overline z^J} K \, , \quad I,J \ne 0 \, , 
\ee
deriving from the K\"ahler potential 
\be
\label{KAHLER}
K = \ln \phi^0 \overline \phi^0 \, ,
\ee
which is defined in terms of the independent physical scalars $z^I$ that are 
\be
z^I = \frac{\phi^I}{\phi^0} \, . 
\ee 
For the ungauged $\cN=2$ Poincar\'e supergravity \eqref{standardSG} there is no scalar potential 
\be
{\cal V}_{\text{no FI}} = 0 \, , 
\ee
and the 
auxiliary fields of the vector and tensor multiplets are dynamically set to zero
\be
\label{scalAUX-0}
X^J_{ij} \equiv 0  
\, , \quad  
F \equiv 0
\, .
\ee
The $SU(2)_{R}$ symmetry gauge connections $\phi_m{}^{ij}$ and the gauge two-form of the tensor multiplet $b_{mn}$
are also auxiliary fields. 
To integrate out $\phi_m{}^{ij}$ it is more convenient to split it into the trace  and traceless parts 
\be
\phi_m{}^{ij} = \Psi_m{}^{ij} + \frac12 \delta^{ij} \hat \phi_m \, , \quad  \Psi_m{}^{ij} \delta_{ij} = 0 \, . 
\ee
The relevant bosonic part of the total Lagrangian \eqref{standardSG} then reads 
\be
e^{-1} {\cal L}_{\phi,\tilde h} = \Psi_m{}^{ij} \Psi^m{}_{ij} - \frac12 \tilde h^a \tilde h_a 
- \tilde h^a \hat \phi_a 
\, .    
\label{ACTphiH}
\ee
The equations of motion of $\phi_m{}^{ij}$ and $b_{mn}$ identically set $\Psi_m{}^{ij} \equiv 0$
while $\hat \phi_m$ and $b_{mn}$ are set to be pure gauge, which we then gauge fix to zero.
As a result also these auxiliary fields are all identically set to zero, 
that is 
\be
\quad \Psi_m{}^{ij} = 0 \, ,
\quad \hat \phi_m = 0 \, , 
\quad \tilde h_m = 0 \, . 
\ee
Then, on-shell, all the $SU(2)_R$ symmetry stops to be gauged and the gravitini are uncharged under the vector multiplets 
$U(1)$s.
Finally we integrate out the real rank-two antisymmetric tensor auxiliary field $W_{ab}$ 
which gives 
\be
\delta W_{\alpha \beta} \quad \Longrightarrow \quad W_{\alpha \beta} 
= 2 \frac{N_{IJ} \overline \phi^J}{N_{KL} \overline \phi^K \overline \phi^L} F^I_{\alpha \beta} \, ,
\label{varW}
\ee
together with its complex conjugate.
The kinetic terms of the vectors then read 
\be
\label{MAXWELL}
e^{-1} {\cal L}_{\text{Maxwell}} = \frac{1}{2} \, \text{Re} \omega_{IJ} \, F^I_{mn} F^{Jmn} 
- \frac{1}{4} \, \text{Im} \omega_{IJ} \, F^I_{mn} F^{J}_{kl} \ve^{mnkl} \, , \quad I = (0,A) \, ,    
\ee
where 
\be
\omega_{IJ} = 2 {\cal F}_{IJ} - 2 \frac{N_{IK} \overline \phi^K N_{JL} \overline \phi^L}{N_{MN} \overline \phi^N \overline \phi^M} \, . 
\ee
Here the field-strength of the graviphoton, that belongs to the supergravity multiplet, is $F^{0}_{mn}$ 
and the field-strengths of the physical vectors are $F^A_{mn}$. 
This concludes the standard derivation of ungauged $\cN=2$ Poincar\'e supergravity from an off-shell setting.
Note that, due to the absence of any scalar potential, 
the vacuum of the previous ungauged $\cN=2$ Poincar\'e supergravity 
coupled to a system of Abelian vector multiplets is $\cN=2$ supersymmetric Minkowski.

\subsection{Gauged $\cN=2$ supergravity}

In this subsection we review the standard Fayet--Iliopoulos term and show how it arises from the off-shell coupling between
the vector multiplets and the tensor multiplet compensator. The addition of this coupling to the ungauged supergravity
\eqref{standardSG} leads to gauged $\cN=2$ supergravity where, on-shell, part of the $SU(2)_R$ symmetry group
remains gauged by a combination of $U(1)$s of the vector multiplets under which the gravitini will be charged. 

The ${\cal L}_{\text{standard FI}}$ Lagrangian describing the standard $\N=2$ Fayet--Iliopoulos term
can be obtained from the following conformal superspace chiral action
\begin{align}\label{PsiW2}
S_{\text{standard FI}} = - \int \rd^4x\, \rd^4\q\, \cE\, \Psi \,\xi_I {W}^I  +c.c.
\end{align}
This describes a locally superconformal extension of a $ b_2\wedge\,\xi _I F_2^I\simeq\,h_3\wedge\, \xi _Iv_1^I$ 
topological action, 
where $b_2=\hf \rd x^n\wedge \rd x^mb_{mn}$
 is the gauge two-form of the tensor multiplet compensator described by the chiral 
prepotential $\Psi$ and possessing the component 3-form field strength $h_3=\rd b_2$,
while $F_2^I=\hf \rd x^n\wedge \rd x^mF^I_{mn}=\rd v_1^I$ is the two-form field strength of the vector multiplet ${W}^I$
possessing component gauge one-form $v_1^I=\rd x^mv_m^I$.
The bosonic sector of \eqref{PsiW2}, which is enough for the purpose of our discussion,  is given by 
\bsubeq\label{bos-standard-FI}
\bea
e^{-1} {\cal L}_{\text{standard FI}} 
&=& 
-\frac18 \xi_I  G^{ij} X_{ij} ^I  
- \xi_I F \phi^I 
+ \frac14 \xi_I \ve^{mnpq} b_{mn} F^I_{pq}  
+ c.c. 
\\
&=& 
-\frac18 \xi_I  G^{ij} X_{ij} ^I  
- \xi_I F \phi^I 
+ \xi_I \tilde{h}^av_a^I
+ c.c. 
\label{bos-standard-FI-2}
\eea
\esubeq
The first terms are the analogue of the flat FI terms where $\x^{ij}$ is now given by $\tilde{\x}G^{ij}$
while the other terms arise necessarily due to the presence of the hypermultiplet compensator $\cG^{ij}$.
As we will see shortly, 
the last term describing the bosonic BF coupling between the gauge two-form $b_2$ and the specific combination 
of two-form field strength $\hat{F}_2=\xi _I F_2^I$ is the source of the gauging.
The off-shell Lagrangian of $\cN=2$ gauged supergravity is given by
\be
\label{gaugedSG}
{\cal L}_{\rm gauged} = \left\{ - {\cal L}_{\rm V} + c.c. \right\} 
+ {\cal L}_{\rm L} 
+ {\cal L}_{\text{standard FI}} 
\, ,
\ee
where ${\cal L}_{\rm V} $ and ${\cal L}_{\rm L} $ were given in the previous subsection.
Let us now reconsider the gauge-fixing conditions and integration of the auxiliary fields described in the previous subsection
once the standard Fayet--Iliopoulos terms are turned on in \eqref{gaugedSG}.

It is clear that, since \eqref{bos-standard-FI} does not depend on the $D$, $W_{ab}$, $A_m$, and $\phi_m{}^{ij}$
fields of the Weyl multiplet of conformal supergravity, their equations of motion will be identical to the ones described in the 
previous subsection, more specifically eqs.~\eqref{varD},  \eqref{varA}, and \eqref{varW}. 
Moreover, the variation with respect to $\phi_m{}^{ij}$ will set $\Psi_m{}^{ij}\equiv 0$ and 
$\tilde{h}_{m}\equiv 0$. We will also impose the same gauge conditions of the previous subsection,
eqs.~\eqref{GaugeFix}--\eqref{GaugeFix-2},
 that fix dilatation and $R$-symmetry together with $b_{mn}\equiv 0$, which can be imposed once $h_{mnp}=0$ on-shell.

The standard FI term, however, modifies the variation with respect to auxiliary fields within the vector and tensor multiplets.
As a result the auxiliary fields that will have different equations of motion are 
the gauge field $\hat \f_m$ of the $\widehat{U}(1)_R$ group, 
the auxiliary fields $X^I_{ij}$ of the vector multiplets, and the auxiliary field $F$ of the tensor multiplet compensator.
These are no longer set to zero on-shell, 
and instead acquire a nontrivial dependence upon the physical fields of the vector multiplets\bsubeq
\label{scalAUX-3}\bea
F &=& -2 \, \xi_I \overline \phi^I \, , 
\\
X^I_{ij} &=& - 4 \, N^{IJ} \xi_J \, \delta_{ij} \, ,
\\
\hat \phi_m&=& 2 \xi_I v^I_m \, ,
\label{phi-xv}
\eea
\esubeq
where equation \eqref{phi-xv} arises from the last term of \eqref{ACTphiH} and \eqref{bos-standard-FI},
while $N^{IJ} N_{JK} = \delta^I_K $. 
The addition of the standard FI term then leads to the following important differences compared to the ungauged 
$\cN=2$ supergravity of the previous section:
\begin{itemize}
\item[{i)}] The standard FI term introduces a nontrivial potential for the scalar sector of the theory whose bosonic
Lagrangian becomes
\bea
\label{VST-2}
e^{-1} {\cal L}_{\text{gauged}}
&=& 
\frac12 R 
- \sum_{I,J \ne 0} g_{I \overline J} \, \partial_m z^I \partial^m \overline z^J 
- {\cal V} 
\non\\
&&
+\frac{1}{2} \, \text{Re} \omega_{IJ} \, F^I_{mn} F^{Jmn} 
- \frac{1}{4} \, \text{Im} \omega_{IJ} \, F^I_{mn} F^{J}_{kl} \ve^{mnkl}
\, , 
\eea
with
\bea
\label{oldV}
{\cal V} = {\cal V}_{\text{standard FI}} = - N^{IJ} \xi_I \xi_J - 2 | \xi_I \phi^I |^2 \, .  
\eea
Remember that $N_{IJ}$, and then $N^{IJ}$, have Lorentzian type signature and then $-N^{IJ} \xi_I \xi_J$ can be
both positive and negative depending on the choice of $\xi_I$.

\item[ii)]
Equation \eqref{scalAUX-3} identifies on-shell 
the abelian vectors of the physical multiplets $v_m^A$ together with the graviphoton $v_m^0$, 
weighted by the Fayet--Iliopoulos coupling constants $\xi_I$, 
with the auxiliary gauge field $\hat \phi_m$ that gauges the $\widehat{U}(1)_R$ subgroup of  $SU(2)_{R}$. 
As a result the gravitini become charged under the Abelian $U(1)$s of the propagating vectors
signalling that the $\widehat{U}(1)_R$ subgroup of the $SU(2)_{R}$ is gauged.
Equation \eqref{scalAUX-3} describes the precise embedding of the $U(1)$s in the residual $R$-symmetry.

It is important to stress again that the term responsible for the gauging of the $R$-symmetry is the $b\wedge f\simeq h\wedge v$
term
in eq.~\eqref{bos-standard-FI}. Without such term in the action on-shell we would still have $\hat{\phi}_m\equiv0$ instead of 
\eqref{scalAUX-3}. This fact will play a distinctive role when we look at the new FI term.

\item[iii)] Even though we have not mentioned many details about the fermionic sector of the theory, 
let us discuss here only the gravitini, as the third important impact of the gauging concerns the generation of 
non-vanishing gravitini masses. 
Indeed, 
the standard FI contains also a term of the form 
$\xi_I \phi^I \delta_{ij} \, \overline \psi_c^i \overline \sigma^{cd} \overline \psi_d^j  +c.c.$ 
that introduces a gravitino mass (see e.g. \cite{Cremmer:1984hj,Butter:2012xg}). 
The terms that contribute to the kinetic and mass terms of the gravitini are given by 
\be
\label{GRAVI}
\begin{aligned}
{\cal L}_{\text{Gravitini}} = & ~\frac12 \ve^{mnpq} \overline \psi_m^j \overline \sigma_n \nabla'_p \psi_{qj} 
- \frac12 \ve^{mnpq} \psi_{mj} \sigma_n \nabla'_p \overline \psi_{q}^{j}  
\\
& + \xi_I \phi^I \delta_{ij} \, \overline \psi_c^i \overline \sigma^{cd} \overline \psi_d^j  
+ \xi_I \overline \phi^I \delta^{ij} \, \psi_{ci} \sigma^{cd} \psi_{dj}  \, , 
\end{aligned}
\ee
where because of the gauging we have 
\be
\nabla'_a \psi_{nj}  = {\rm D}_a(\omega) \psi_{nj}  
- \frac12 e_a^{\ m} \hat \phi_m \epsilon_{j k}  \delta^{ki} \psi_{n i} 
- \ri A_a \psi_{nj}  \, , 
\ee
and the gauge fields are given by \eqref{varA} and \eqref{scalAUX-3}. 
There are of course various other terms quadratic in the gravitini, 
however in \eqref{GRAVI} we have included only the ones that contribute to the kinetic terms and to the 
mass.\footnote{The complete action can be found in \cite{Cremmer:1984hj}.} 
Notice, in particular, that the value of the gravitino mass is 
\be
m_{3/2}^2 = |\xi_I \phi^I |^2  \, ,
\ee 
where we stress that we are considering only models of $\cN=2\to\cN=0$ breaking with both local supersymmetries 
broken at the same scale.
Under the $\widehat{U}(1)_R$ gauged subgroup of $SU(2)_{R}$ the two gravitini rotate 
to each other, 
that is 
\be
\widehat{U}(1)_R : \quad \delta \psi_m{}_1 = - \alpha \, \psi_m{}_2 \, , \quad \delta\psi_m{}_2 = \alpha \, \psi_m{}_1 \, , 
\ee
which is indeed a symmetry of the gravitini mass terms, 
and $\alpha$ is a $\widehat{U}(1)_R$ parameter.

\item[iv)] 

Clearly, 
because the gauging generates a scalar potential, 
it should also contribute to the supersymmetry transformations of the gaugini. 
In particular, 
since the gaugini transformations 
have the form  (we neglect terms that will be zero on a Lorentz invariant vacuum)
\be
\label{DDLLL}
\delta \lambda_\alpha^{Ii} = -\frac12 \left( X^{Iij} - 2 \delta^{ij} F \phi^I  \right) \epsilon_{i\alpha}(x) + \dots  
\ee
any vev for $X^{Iij}$ and $F \phi^I$ will generate a shift that will signal a supersymmetry breaking.\footnote{It is convenient in 
our discussion here to include also the compensator gaugini $\lambda^{0j}$, 
even though after gauge fixing they are subject to the condition $N_{IJ} \overline \phi^I \lambda^{Jj} = 0$, 
which is imposed by integrating out the auxiliary fermions $\Sigma_{i\alpha}$.} 
This observation allows us to introduce a consistency check 
for supersymmetry breaking. 
We will illustrate this here 
for the gauged supergravity but it can be also used in the ungauged case, 
and it will be very helpful for the check of self-consistency of the new FI terms that we will introduce later. 
Because in our analysis the scalar potential is generated {\it only} by auxiliary fields (even when we include new FI terms), 
it means that it will have the form 
\be
\label{VXFon}
{\cal V}_{\text{On-shell}} = - \frac12 \left(  \frac{1}{16} N_{IJ} X^I_{ij} X^{J ij}  + |F|^2 \right) \Big{|}_{X^{ij}\, , 
F\,   {\text{are evaluated on-shell}}} \, . 
\ee  
Once more, notice that, due to the Lorentzian type of signature of $N_{IJ}$, the first term in
${\cal V}_{\text{On-shell}}$ is not negative definite and allows in principle for both positive and negative 
dynamically generated cosmological constants, whereas the $F$ auxiliary superfield in 
\eqref{VXFon} always leads to a universally negative contribution to the potential.
For gauged supergravity, the values of the auxiliary fields \eqref{scalAUX-3} are inserted in \eqref{VXFon}. 
From the gaugini supersymmetry transformations we see that supersymmetry restoration means that we have 
\be
\label{nobreak}
\langle \delta \lambda^{Ii} \rangle = 0  \quad \Longrightarrow \quad X^{Iij} = 2 \delta^{ij} F \phi^I \, . 
\ee
Therefore when supersymmetry is restored the condition \eqref{nobreak} will hold for the vacuum 
and the scalar potential \eqref{VXFon} will have the vacuum value 
\be
\label{susyVAC}
\langle {\cal V}_{\text{SUSY}} \rangle = - \frac34 |\langle F \rangle|^2 = - 3 m_{3/2}^2 \, . 
\ee 
This expression is the standard expression that 
relates the gravitino mass to the vacuum energy for supersymmetric anti-de Sitter supergravity. 
As a result, 
when we have a vacuum that satisfies \eqref{susyVAC} 
we will know that supersymmetry may be preserved. 
More importantly, however, 
when we have a vacuum that violates \eqref{susyVAC} we will know that 
supersymmetry is definitely broken. 
This happens because supersymmetric vacua always satisfy \eqref{susyVAC}. 
On the contrary, de Sitter vacua, that will be the main focus of our analysis here, 
will always violate \eqref{susyVAC} and therefore guarantee the spontaneous breaking of supersymmetry.

\end{itemize}

The presence of the standard FI term leads to a very rich dynamics and structure of vacua, including AdS and dS,
see e.g. 
\cite{Cremmer:1984hj,Catino:2013syn,DAuria:1990qxt,Andrianopoli:1996vr,Andrianopoli:1996cm,DallAgata:2003sjo,FVP,Trigiante:2016mnt}. 
However, 
the standard FI term is incompatible with 
supergravity-matter systems that include also physical charged hypermultiplets
 (see for example 
 \cite{DAuria:1990qxt,Andrianopoli:1996vr,Andrianopoli:1996cm,DallAgata:2003sjo,VanProeyen:2004xt,FVP,Trigiante:2016mnt} 
 and \cite{Bagger:1987rc} for an off-shell superspace derivation of this no-go theorem). 
We expect that the latter limitation can
naturally be overcome when new FI terms are added to ungauged $\cN=2$ supergravity, 
but we will not study such extension in this article, 
rather we will only work with physical vector multiplets.

Before closing this section let us return to the formula  \eqref{noDS} that we presented in the introduction and study 
it within a model with a single physical vector multiplet ${W}^1$. 
To contrast the standard FI to the new one that we will introduce later, 
we will switch-on only the  FI term parameter for the ${W}^1$. 
For clarity we will study explicitly the ${\mathbb C}P^{1}$ model with 
\be
\label{CCPP11}
{\cal F} = \frac14 ( \phi^0 )^2 - \frac14 ( \phi^1 )^2 \, , 
\ee
which gives 
\be
N_{I J} = \begin{pmatrix}
   1  & 0  \\
   0  & - 1    
  \end{pmatrix} \, . 
\ee
We define $z = \phi^1 / \phi^0$, 
and we find from \eqref{GaugeFix} 
\be
\phi^0 \overline \phi^0 = \frac{1}{1 - |z|^2} \, , 
\ee
therefore the K\"ahler potential and the K\"ahler metric take the form 
\be
\label{KgCP1}
K = - \ln \left( 1 - |z|^2  \right) \, , \quad g_{z \overline z} = \frac{1}{(1 - |z|^2)^2} . 
\ee
Notice that the moduli space is bounded by $|z|^2 < 1$. 
Now we switch on only the FI term for the physical vector mutiplet therefore we fix the $\xi^I$ to have the form 
\be
\xi^I =  ( 0 , \xi ) \, , \quad  \xi \in \mathbb{R} \, , 
\ee
which brings the scalar potential to the form 
\be
\label{SPphys}
{\cal V}(z, \overline z) = 
\xi^2 \left( 
1 - 2 \frac{|z |^2}{1 - |z|^2} \right) \, . 
\ee
An inspection of the scalar potential \eqref{SPphys} shows that there is {\it no} critical point (de Sitter or anti-de Sitter) 
except  for the $z=0$ which is de Sitter and unstable, thus verifying \eqref{noDS}.

The situation changes when we switch on also the FI term for the compensator vector multiplet setting $\xi^0 \ne 0$. 
An anti-de Sitter supersymmetric (thus stable) critical point arises, but the de Sitter critical point is still unstable, 
in agreement with \eqref{noDS}.

\section{New deformations of $\cN=2$ supergravity}  
\label{NewFI-sugra}

In this section we introduce the new FI terms and the uplift terms of the $\cN=2$ supergravity 
utilizing a composite goldstino built from physical vector multiplets. 
Besides making some of the analysis technically easier, the composite goldstino clearly indicates how supersymmetry breaking 
is sourced by the mediating vector multiplet. 
Finally we also study the new scalar potentials and focus on the construction of models admitting stable de Sitter vacua.

\subsection{A composite goldstino}
\label{composite-goldstino}

In the flat case we have shown how given a vector multiplet that mediates $\cN=2\to\cN=0$ supersymmetry breaking
the goldstino fields are related to the gaugini according to \eqref{flat-composite-goldstino}.
The aim of this subsection is to lift the same result to supergravity. 
Before entering into the details of the construction, which is technically more involved and richer than 
the one of section \ref{section-2},
it is worth anticipating the results that ultimately closely resemble the flat case. 
We will show that in the curved case the component goldstini fields satisfy 
\be
\g_{\alpha i} 
=
 - 4 \frac{\cX_{ij}}{\cX^{kl}\cX_{kl}} \l_\a^{j}
 +\dots
 ~,
\label{curvedGammaLambda-3}
\ee
where $\cX^{ij}$ are a curved extension of the vector multiplet auxiliary fields in \eqref{Xflat} and, 
as we will discuss soon in more detail,
include contributions depending on the supergravity compensators. 
Importantly, $\cX^{ij}$ is the field appearing in the Poincar\'e supergravity
supersymmetry transformations of the gaugini 
\be
\label{dLdX} 
\delta \lambda_\alpha^{i} = -\frac12 {\cal X}^{ij} \epsilon_{\alpha j}(x) + \dots 
\ee 
where the goldstini \eqref{curvedGammaLambda-3} transform as a shift
\be
\delta \gamma_i = \epsilon_i(x) + \dots  
\label{goldstini-curved-shift}
\ee 
Then, as in the flat case, supersymmetry is completely broken when 
\be
\label{Xvevs} 
\langle {\cal X}^{ij} {\cal X}_{ij} \rangle \ne 0 \, . 
\ee
Let us now dig into the technical derivation and analysis of the previous results.

We consider an  $\cN=2$ Abelian vector multiplet coupled to conformal supergravity which is 
described by the  superfield strength $W$ satisfying the constraints
\eqref{DEFW} and \eqref{DWYW}.
The construction of a composite goldstino does not require a priori 
to impose $W$ to be a restricted $\cN=2$ chiral multiplet, as we have done in \eqref{DEFW}. 
In a generalization one can indeed relax the second condition in \eqref{DEFW}
and $W$ might be replaced by a function of other multiplets. 
In fact, it is simple to realize that the construction below only relies on the existence of some multiplet that mediates supersymmetry 
breaking with a fermionic field working as a goldstino. 
In any case, for simplicity and clarity, in our paper we will only focus on the case of a single vector multiplet mediating 
supersymmetry breaking.

We remind that 
the component fields of the vector multiplet $\phi$, $\lambda^i_\alpha$, and  $X^{ij}$, were defined in \eqref{CompVect}.
To construct a composite goldstino, 
we will assume that both supersymmetries are broken 
spontaneously by the auxiliary fields $X^{ij}$ of $W$ 
and therefore the gaugini $\lambda_\alpha^{i}$, will serve as the goldstini. 
It is important to stress a difference between the $\cN=1$ and $\cN=2$ case. 
For an $\cN=1$ vector multiplet coupled to conformal supergravity the gaugino is a primary field. In the $\cN=2$ case this is 
not the 
case, in fact, under $S$-supersymmetry, 
it holds $S^\a_i\l_\b^{j }\propto \phi$. This makes the construction of a composite goldstino multiplet 
more involved and, in particular, it implies
that we need to introduce compensators in the $\cN=2$ analysis.

We will now lift formula \eqref{Xsusy} to supergravity and construct 
a primary nilpotent chiral superfield $X$ which will be a composite of the $\cN=2$ vector multiplet $W$. 
To do so, we first need to assume the existence of a real conformal compensator superfield $C$ 
defined to be a primary ($K^AC=0$) such that
\be
\mathbb{D} \, C = 2C \, , \quad Y \, C =0 \, ,\quad
C\ne0~.
\ee
We will also assume the existence of a complex 
compensator for $U(1)_R$, that we will denote $Z$ (not necessarily related to $C$ nor $W$), such that
\be
\mathbb{D} \, Z = Z \, , \quad Y \, Z =-2Z \, ,\quad
Z\ne0~.
\ee
Note that in general $C$ and $Z$ might be non-trivial composite superfields of other compensators (e.g. ${W}^0$ and ${\cal G}_{ij}$), 
as we will indeed set later. 
However, for the scope of this subsection we only need their existence. 
Their main use is to turn $\de_\a^i W$ and $X$ into primary superfields.

It was shown in \cite{Butter:2012xg} that by using a real conformal primary compensator of dimension 2,
as $C$,
it is possible to construct operators $\cD_A$
 which are completely
\emph{inert} under dilatation, conformal, and $S$-supersymmetry
transformations. The new covariant derivatives are given by
\begin{subequations}
\label{Dcov}
\begin{align}
\cD_\alpha^i &= e^{-U/4}
	\left(\nabla_\alpha^i
	- \nabla^{\beta i} U M_{\beta \alpha}
	+ \frac{1}{4} \nabla_\alpha^i U Y
	- \nabla_\alpha^j U J_j{}^i\right)~, \\
\overline\cD^\dalpha_i &= e^{-U/4}
	\left(\overline\nabla^\dalpha_i
	+ \overline\nabla_{\dbeta i} U \overline M^{\dbeta \dalpha}
	- \frac{1}{4} \overline\nabla^\dalpha_i U Y
	+ \overline\nabla^\dalpha_j U J^j{}_i \right)~,
\end{align}
\end{subequations}
where $U := \log C$. 
These derivatives, whose algebra is given in appendix \ref{AppendixConfSuperspace} in eq.~\eqref{algebraCalD},
are such that if $T$ is some conformally primary
tensor superfield of vanishing dilatation weight, then $\cD_\alpha^i T$
and $\overline\cD^\dalpha_i T$ are as well.

Given a vector multiplet $W$ coupled to conformal supergravity, it is then useful to introduce the dimension zero 
primary superfield
\bea
\cW:=C^{-1/2}W~,~~~~~~
\mathbb D\cW=0~,~~~
Y\cW=-2\cW
~.
\eea
This is chiral with respect to the $\cD_A$ covariant derivatives
\bea
\cDB^\ad_i\cW=0=\cD_\a^i\overline{\cW}
~,
\label{chiral2}
\eea
and satisfies the Bianchi identity
\bea
(\cD^{ij}+4S^{ij})\cW
=(\cDB^{ij}+4\overline S^{ij})\overline{\cW}
~,~~~~~~
\cD^{ij}:=\cD^{\a(i}\cD_\a^{j)}
~,~~~
\cDB_{ij}:=\cDB_{\ad(i}\cDB^\a_{j)}
~,
\label{BI2}
\eea
with $S^{ij}$ and $\overline S^{ij}$ being some of the torsion components appearing in the algebra of covariant derivatives
$\cD_A$; see appendix \ref{AppendixConfSuperspace} and in particular \eqref{dim-1-torsions}.
The results \eqref{chiral2} and \eqref{BI2} are direct consequence of \eqref{DEFW}.
By using the $\cD_A$ derivatives we define the descendant spinor  $\L_\a^i$ and its complex conjugate $\overline{\L}^\ad_i$ as
\bsubeq\label{primary_gaugino}
\bea
\L_\a^i
&\equiv&
\cD_\a^i\cW|
=
C^{-3/4}
	\left[\nabla_\alpha^i W
	-  (\nabla_\alpha^i \log C) W
\right]|
~,
\\
\overline\L^\ad_i
&\equiv&
\cDB^\ad_i\overline{\cW} |
=
C^{-3/4}
\left[\deb^\a_i \overline{W} 
-  (\deb^\ad_i \log C) \overline{W} 
\right]|
~,
\eea
\esubeq
which is such that
\bea
K^A\L_\a^i=\mathbb D\L_\a^i=0~,~~~Y\L_\a^i=-\L_\a^i
~.
\eea
The field $\L_\a^i$ is a primary extension of the gaugino $\l_\a^{i}$. 
Below we will indicate with $\L_\a^i$ the superfield $\cD_\a^i\cW$
and it will be clear from the context if we refer to the superfield or its lowest component.

Now that we have introduced the previous technical ingredients, we can define
\be
\label{Xsugra}
X = 
Z^{-2}\overline \Delta |\cD \cW|^8
\, , 
\ee 
where $\overline \Delta = \frac{1}{48}  \overline \nabla^{ij} \overline \nabla_{ij}$ is the chiral 
projector\footnote{
The nomenclature 
``projector'' is misleading, 
since $\overline \Delta\overline \Delta=0\ne\overline \Delta$ 
which follows from $\deb^\ad_i\overline \Delta \equiv0$, but it is conventionally used, and we will follow this convention here. 
Note that, 
given an arbitrary superfield $U(z)$, the superfield $\Phi(z):=\overline \Delta U(z)$ is by construction chiral,
$\deb^\ad_i\F=0$. For the construction
of the chiral projector in $\cN=2$ curved superspaces, besides \cite{Butter4DN=2}, 
see \cite{Muller:1989uhj} and the more recent normal coordinates analysis of \cite{Kuzenko:2008ry}.
} 
in conformal superspace \cite{Butter4DN=2} and we have defined 
\bsubeq
\bea
&(\cD\cW)^{ij}:=\cD^{\a i}\cW\cD_\a^j\cW
~,~~~
(\cDB\overline\cW)_{ij}:=\cDB_{\ad i}\overline\cW\cDB^\ad_j\overline\cW
~,\\
&(\cD\cW)^4:=\frac{1}{3}(\cD\cW)^{ij}(\cD\cW)_{ij}
~,~~~
(\cDB\overline\cW)^4:=\frac{1}{3}(\cDB\overline\cW)^{ij}(\cDB\overline\cW)_{ij}
~,\\
&|\cD\cW|^8:=(\cD\cW)^4(\cDB\overline\cW)^4
~. 
\eea
\esubeq
The scalar superfield $X$ is by construction chiral ($\overline \nabla_i^{\dot \alpha} X  = 0$) 
and primary ($K^AX=0$), 
it reduces to \eqref{Xsusy} in the global limit, 
and its weights are 
\be
\mathbb{D} \, X = 0  \, , \quad Y \, X = 0 \, . 
\ee 
By construction it also satisfies $\cDB_i^{\dot \alpha} X  = 0$.
Moreover, by using arguments similar to the ones used in appendix \ref{Proof-Nilpotency}, 
that easily extend to the supergravity case,
one can show that the superfield $|\cD\cW|^8$ satisfies the following nilpotency conditions 
\be
\de_{A_1}\de_{A_2}\cdots\de_{A_m}|\cD\cW|^8\,\de_{B_1}\de_{B_2}\cdots\de_{B_n}|\cD\cW|^8=0
~,~~~~~~
\forall~
m,n=0,1,\cdots,7~,~~~
m+n\leq 7
~,
\label{nilp-1-sugra}
\ee
together with the following expressions containing eight covariant derivatives 
\bea
&&\de_C\de_{A_1}\cdots\de_{A_m}|\cD\cW|^8
\,
\de_{A_{m+1}}\cdots\de_{A_7}|\cD\cW|^8=
\non\\
&&~~~~~~=
-(-)^{\ve(C)\big(\sum_{k=1}^m\ve(A_k)\big)}
\de_{A_1}\cdots\de_{A_m}|\cD\cW|^8
\,
\de_C\de_{A_{m+1}}\cdots\de_{A_7}|\cD\cW|^8
~,
\label{nilp-2-sugra}
\eea
that holds $\forall~m=0,1,\cdots,7$. In particular, it holds
\bea
\Delta |\cD\cW|^8\overline \Delta |\cD\cW|^8
= |\cD\cW|^8
\overline \Delta\Delta |\cD\cW|^8
= |\cD\cW|^8
\Delta\overline \Delta |\cD\cW|^8
~,
\label{nilp-3-sugra}
\eea
which will be useful later. All these conditions hold also when using $\cD_A$ derivatives instead of the $\de_A$ ones
and, exactly as in appendix \ref{Proof-Nilpotency}, simply derive from the fact that the product of more than
four $\cD_\a^i\cW$ is identically zero (the same holds for a product of more than four $\cDB^\ad_i\overline\cW$).

Due to the aforementioned results, 
the composite superfield $X$ of eq.~\eqref{Xsugra}
satisfies a series of covariant nilpotency conditions that have been presented in \cite{Kuzenko:2017zla}, 
which are 
\be
X^2 = 0 \, , \quad X \, \nabla_A \nabla_B X = 0 \, , \quad X \, \nabla_A \nabla_B \nabla_C X = 0 \, . 
\label{nilpotency-curved-1}
\ee 
As described in details in \cite{Kuzenko:2017zla}, 
if
\be
\langle \D X\rangle\ne0 
\label{DXvev}
\ee 
is satisfied then supersymmetry is completely broken
and the $\cN=2\to\cN=0$ goldstino multiplet is described by
\be
\label{Gsugra}
\Gamma_{\alpha i} = - \frac{1}{12} \frac{\nabla^{j}_\alpha \nabla_{ij} X }{\Delta X} \, ,
\ee
which generalizes \eqref{Gsusy}.
The way the superspace derivatives act on $\Gamma_{\alpha i}$ is presented in formulas (4.8a) and (4.8b) of  \cite{Kuzenko:2017zla}, 
which is essentially the curved superspace generalization of \eqref{NSWrepr}. 
For convenience we can repeat here the main properties of $\G_{\a i}$.
It holds
\bsubeq\label{deXi-sugra}
\bea
\de_\b^j\G^\a_{ i}
&=&
\d_\b^\a\d^j_i
~,
\\
\deb^\bd_{ j}\G^\a_{i}
&=&
2\ri \G_{\g j}\de^{\g\bd}\G^\a_{ i}
-\ri\ve_{ij}(\de^\a{}_{\gd} \overline{W}^{\bd\gd})\G^4
-\ri\ve_{ij}\overline{W}^{\bd\gd}\de^\a{}_{\gd}\G^4
\non\\
&&
-2\ri\ve_{ij}\overline{W}^{\bd\gd} \G_{kl}\G^{\a k}\de_{\g\gd}\G^{\g l}
-\frac{4\ri}{3}\overline{W}^{\bd\gd} \G_{k(i} \G^{\a k}\de_{\g\gd}\G^\g_{j)}
\non\\
&&
-\frac{1}{3}\ve_{ij}(\deb_{\gd}^{k} \overline{W}^{\bd\gd} )\G_{kl}\G^{\a l}
-\frac{2}{3}(\deb_{\gd(i} \overline{W}^{\bd\gd} )\G_{j)k}\G^{\a k}
~,
\\
S^\b_j\G^\a_i
&=&
2\ve^{\a\b}\G_{ij}
+2\ve_{ij}\G^{\a\b}
~,~~~
\overline{S}_\bd^j\G^\a_i=0
~,~~~
K^a\G_{\a i}=0
~.
\label{SGamma}
\eea
\esubeq
Note that $\G_{\a i}$ is not a primary\footnote{Though not necessary for our analysis, a primary 
extension of $\Gamma_{\alpha i}$ can be straightforwardly constructed by using the compensator $Z$
along the same line of the results presented in \cite{Kuzenko:2017zla}.}
 but remarkably, thanks to \eqref{SGamma},
the superfield $\G^4=\frac{1}{3} \, \Gamma^{ij} \Gamma_{ij} =-\frac{1}{3} \, \Gamma^{\a\b} \Gamma_{\a\b}$ 
turn out to be a primary such that
\bea
K^A\G^4=0
~,~~~~~~
\mathbb D \G^4=-2\G^4
~,~~~~~~
Y\G^4=-4\G^4
~.
\eea
An important property of $\G^4$ is
\bea
\deb^\ad_{ i}\G^4
&=&
-2\ri\G^4\de^{\g\ad}\G_{\g i}
~.
\label{debXi^4}
\eea
This relation can be  used to check  that 
\be
X=\G^4\D X
~,
\label{XGammaDX}
\ee 
is chiral. 
In fact, the equation \eqref{XGammaDX} solves the constraints \eqref{nilpotency-curved-1}. 
Another useful relation that derives from \eqref{nilp-1-sugra}--\eqref{nilp-3-sugra}, 
that extends the flat relation \eqref{XcalX}, is
\be
\begin{aligned}
X \overline{X} 
& = 
\frac{1}{Z^{2}\overline{Z}^2}
\overline \Delta |\cD\cW|^8\,
\Delta |\cD\cW|^8
= 
\frac{1}{Z^{2}\overline{Z}^2}
 |\cD\cW|^8
\overline \Delta\Delta |\cD\cW|^8
= 
Z^{-2}
 |\cD\cW|^8
\overline \Delta\overline{X}
~.
\end{aligned}
\ee
By dividing the previous expression by $\Delta X\overline\Delta \overline X$, 
and using again the nilpotency relations \eqref{nilp-1-sugra}--\eqref{nilp-3-sugra}, 
one gets
\be
|\G|^8:=\G^4\overline{\G}^4=\frac{X \overline{X} }{\Delta X\overline\Delta \overline X}
= 
\frac{Z^{-2} |\cD\cW|^8}{{\Delta X}}
= 
\frac{Z^{-2} |\cD\cW|^8}{{\Delta Z^{-2}\overline \Delta |\cD\cW|^8}}
= 
\frac{|\cD\cW|^8}{{\Delta \overline \Delta |\cD\cW|^8}}
~,
\label{GLsugra}
\ee
which extends \eqref{ident} to the supergravity case and shows explicitly how $|\G|^8$ is expressed in terms of the primary gaugini. 
Thanks to nilpotency, one can also derive another equivalent form of $|\G|^8$ that will be useful soon, 
that is 
\be
|\G|^8
= 
\frac{C^{-2}|\cD\cW|^8}{{\cD^4\overline \cD^4 |\cD\cW|^8}}
~,
~~~~~~
\cD^4:=\frac{1}{48}\cD^{ij}\cD_{ij}
~,~~~
\overline\cD^4:=\frac{1}{48}\overline\cD^{ij}\overline\cD_{ij}
~.
\label{GLsugra-2}
\ee

Let us turn back to the self consistency of the previous construction.
It is clear that for the existence of the composite goldstino $\G_{\a i}$, and hence $\cN=2\to\cN=0$ local supersymmetry breaking,
a necessary condition to be satisfied is eq.~\eqref{DXvev} which is equivalent to the condition that the bosonic part of 
the denominators of \eqref{GLsugra} and \eqref{GLsugra-2} have nonzero vev 
\be
\langle \Delta \overline \Delta |\cD\cW|^8\rangle \ne  0 
~,\quad
\Longleftrightarrow\quad
\langle \cD^4\overline \cD^4 |\cD\cW|^8\rangle \ne  0 
~.
\label{SUSY-breaking-3}
\ee
In the flat case these are identically satisfied once $\langle X^{ij}X_{ij}\rangle \ne  0 $.
In the supergravity case, as already mentioned, due to the presence of the compensators, the situation is more subtle. 
To investigate this issue one can compute the purely bosonic part of $\Delta \overline \Delta |\cD\cW|^8$
and equivalently of $\cD^4 \overline \cD^4 |\cD\cW|^8$.
By purely dimensional grounds, and by the requirement that all eight fermions $\cD_\a^i\cW$ and $\cDB^\ad_i\overline{\cW}$ 
of eq.~\eqref{primary_gaugino} in $|\cD\cW|^8$ are saturated by the eight spinor derivatives in $\Delta \overline \Delta$,
it is clear that the bosonic part of $\Delta \overline \Delta |\cD\cW|^8$ is given by an eighth order product of terms such as 
$\de_{ij}W=\deb_{ij}\overline{W} $, 
 $\de_aW$ and $\de_a\overline{W}$, the vector field strengths $F_{\a\b}\propto\de^k_{(\a}\de_{\b)k}W$ and 
 $\overline{F}^{\ad\bd}\propto\overline\de_k^{(\ad}\overline\de^{\bd)k}\overline{W}$,
 but will also depend on  the supergravity compensator $C$ in combinations given by the superfields $S^{ij}$, $\overline{S}^{ij}$,
 $G_{\a\ad}$, $G_{\a\ad}{}^{ij}$, $X_{\a\b}$, $\overline{X}_{\ad\bd}$
defined in \eqref{dim-1-torsions}.\footnote{The dependence upon the super-Weyl tensor, $W_{\a\b}$ and $\overline{W}_{\ad\bd}$, 
appears only at higher orders in fermions.}
Actually, it is simpler to understand the dependence of the bosonic part of $\cD^4 \overline \cD^4 |\cD\cW|^8$.
This clearly depends only on eighth order combinations of 
$\cD^{ij}\cW$, $\cDB^{ij}\overline\cW$, $\cD^k_{(\a}\cD_{\b)k}\cW$, $\cDB^{(\ad}_k\cDB^{\bd)k}\overline\cW$,
and $\cDB^\ad_i\cD_\b^j\cW=\big(-2\ri\d_i^j\cD_\b{}^\ad\cW+4\d_i^jG_\b{}^\ad\cW+4\ri G_\b{}^\ad{}_i^j\cW\big)$
and its complex conjugate. 
Assuming that the vacua preserve 4D Lorentz invariance the vev of $\langle\cD^4 \overline \cD^4 |\cD\cW|^8\rangle$ 
can only be a function of $\langle \cX_{ij}\rangle$ and $\langle \overline\cX_{ij}\rangle$
where 
\bsubeq
\bea
\cX_{ij}
&:=&
\cD_{ij}\cW|
=\big[
(C^{-1}|)X_{ij} 
-4 (C^{-1/2}S_{ij}|)\,\phi\big]
~,
\\
\overline{\cX}_{ij} 
&:=&
\cDB_{ij}\overline\cW|
=\big[
(C^{-1}|) X_{ij} 
-4 (C^{-1/2}\overline S_{ij}|) \,\overline \phi
\big]
~.
\eea
\esubeq
Up to fermion terms, and neglecting terms other than  $\cX_{ij}$ and $\overline\cX_{ij}$ (the full expression will be given elsewhere),
a simple calculation shows that it holds 
\bsubeq
\bea
&&
\cD^4(\cD\cW)^4=\frac{1}{64}[(\cD^{ij}\cW)(\cD_{ij}\cW)]^2+\dots
~,~~~~~~
\cDB^4(\cDB\overline\cW)^4=\frac{1}{64}[(\cDB^{ij}\overline\cW)(\cDB_{ij}\overline\cW)]^2+\dots
~,~~~~~~
\\
&&
\cD^4 \overline \cD^4 |\cD\cW|^8
=
\cD^4(\cD\cW)^4\, \overline \cD^4 (\cDB\overline{\cW})^4
+\dots
=\frac{1}{64^2}|(\cD^{ij}\cW)(\cD_{ij}\cW)|^4
+\dots
~,
\eea
\esubeq
which implies
\bea
\langle\cD^4 \overline \cD^4 |\cD\cW|^8\rangle| 
=\frac{1}{64^2}\langle|\cX^{ij}\cX_{ij}|^4\rangle
+\dots
~.
\eea

As already mentioned, in this paper we will always assume for simplicity that $\langle X^{ij} X_{ij} \rangle\ne0$
implies $\langle \cX^{ij}\cX_{ij}\rangle\ne0$ so that supersymmetry is completely broken, 
and we will {\it a posteriori} cross-check that this assumption is valid as we have explained in the previous section. 
In our examples in the next sections we will focus on the cases where $W$ is a physical vector multiplet, 
because it will be utilized for the new FI terms.

The situation is more subtle if $W$ is chosen to be a compensator  used to describe Poincar\'e supergravity. 
In this case $\Delta X$ is a function of purely geometric tensors and on a background whose vacuum preserves 
some supersymmetry we should have $\langle\Delta X\rangle\equiv 0$, 
in accordance with our earlier discussions. 
Although a complete analysis of this problem is beyond the scope of our paper, we can check this property for the 
simplest nontrivial $\cN=2$ supersymmetric background -- 4D anti-de Sitter (AdS) -- which is the vacuum, e.g., of pure gauged 
supergravity without physical vector multiplets and with a cosmological constant term given by a standard FI term for the 
vector multiplet compensator $W=Z={W}^0$. 
Assuming that the dilatation compensator is $C=\cG$,
a straightforward calculation, along the line given for the  gauged ${\mathbb C}P^{1}$ model of the previous section, 
 shows that on-shell it holds identically
$\cX^{(W_0)}_{ij}=(X_{ij}^{({W}_0)}-4S_{ij}|_{\q=0})\equiv0$, as expected.\footnote{Note that choosing $C=\cG$
one can show that $S^{ij}|= 
\frac{1}{2}G^{-5/2}G^{ij}\Big(F-\hf G^{-2} G_{kl}\chi^{kl}\Big)$ which we will use later.} 
The same statement can be derived directly in superspace
by looking at the superspace equations of motion given in \cite{Butter:2011ym}
where it was shown that $\cN=2$ AdS$_4$ superspace is a solution of pure $\cN=2$ AdS supergravity.
The fact that for this model we find $\cX_{ij}^{({W}^0)}=0$ is actually also quite intuitive.
As described in detail in \cite{Kuzenko:2008qw}, $\cN=2$ AdS$_4$ superspace is characterized by the presence of a so-called
intrinsic vector multiplet described by a field strength superfield $\widetilde{{\cW}}$ that is covariantly constant, 
that is $\bm \cD_A\widetilde{{\cW}}=0$, where the derivatives $\bm \cD_A$ are the $\cD_A$ derivatives 
evaluated on the AdS$_4$ solution. 
In an appropriate gauge, the intrinsic vector multiplet arises as the on-shell value of the vector multiplet compensator of the 
off-shell pure AdS supergravity of \cite{Butter:2011ym}. 
By construction then it is clear that $\langle\cX_{ij}^{({W}^0)}\rangle=\bm \cD^{ij}\widetilde{{\cW}}=0$.

Let us turn back to the general case and consider again \eqref{Gsugra} to see how the goldstini are related to the gaugini. 
If we focus only on contributions linear in fermions, 
by using arguments similar to the ones used above,
the following factorization holds 
\be
\Gamma_{\alpha i} 
=
 - \frac{C^{-1/4}}{12} \frac{\cD^{j}_\alpha
  \cD_{ij}(\cD\cW)^4\cDB^4 (\cDB\overline\cW)^4}{\cD^4(\cD\cW)^4\cDB^4 (\cDB\overline\cW)^4} 
 +\dots
=
 - \frac{C^{-1/4}}{12} \frac{\cD^{j}_\alpha \cD_{ij}(\cD\cW)^4}{\cD^4(\cD\cW)^4} 
 +\dots
 ~.
\ee
If again we restrict only  to terms depending on $\cX^{ij}$ and $\overline\cX^{ij}$, 
it is simple to prove the following result
\be
\cD^{j}_\alpha \cD_{ij} (\cD\cW)^4
=
\frac{3}{4}(\cD^{kl}\cW)(\cD_{kl}\cW)(\cD_{ij}\cW)\cD_\a^{j}\cW
+\dots
~,
\ee
leading to
\be
\Gamma_{\alpha i} 
=
 - 4C^{-1/4} \frac{(\cD_{ij}\cW)}{(\cD^{kl}\cW)(\cD_{kl}\cW)} \cD_\a^{j}\cW
 +\dots
 ~.
\label{curvedGammaLambda-2}
\ee
Once more, note that, compared to the flat case, 
due to \eqref{primary_gaugino}, there is also a dependence on 
$\de_\a^iC$ and $\deb^\ad_iC$ which will eventually simplify upon taking an appropriate conformal gauge fixing $C=1$. 
Indeed, 
in our examples we will have $C={\cal G}$ and in this case the gauge fixing is $G=1$, 
therefore, as already anticipated, the component goldstini fields satisfy eq.~\eqref{curvedGammaLambda-3}.
Despite the complicated form of the precise expression, it is important to stress that, in this setup, in general
\eqref{curvedGammaLambda-3} can be inverted to express the $\lambda^{j }$ in terms of the goldstini
thanks to the non-vanishing vevs of $X^{kl} X_{kl} $ and $\cX^{kl}\cX_{kl}$. 
Then, the supersymmetry transformations of the gaugini 
\eqref{dLdX}  imply  that the $\gamma_i$, given by \eqref{curvedGammaLambda-3}, transform as a shift, 
that is eq.~\eqref{goldstini-curved-shift}.
Finally in agreement with our earlier discussion supersymmetry is completely broken when 
$\langle {\cal X}^{ij} {\cal X}_{ij} \rangle \ne 0$ \eqref{Xvevs}. 
In the next subsection where we study the $\Gamma_{\alpha i}$ with explicit compensators we will see 
how exactly \eqref{dLdX} is related to \eqref{DDLLL}.

To recap, 
we have seen that it is always possible to construct a composite $\cN =2$ goldstino of the type studied in \cite{Kuzenko:2017zla} 
within supergravity, 
by employing a reduced chiral $\cN=2$ (vector multiplet field strength) superfield. 
The self-consistency of such construction requires however that the auxiliary field components of $W$ acquire a vev. 
We will see now how a new type of Fayet--Iliopoulos term can be introduced with the use of the composite goldstini, 
in such a way that it will also guarantee the self-consistency of the construction.

\subsection{New Fayet--Iliopoulos terms in ungauged supergravity}
\label{section-new-FI}

We are now in position to present the new $\cN=2$ Fayet--Iliopoulos term in supergravity. 
We will construct such a term for a 
vector multiplet superfield $W$ defined in \eqref{DEFW} and \eqref{DWYW} assuming a priori that 
the conditions \eqref{Xvevs}, \eqref{DXvev}, and \eqref{SUSY-breaking-3}  
for the complete breaking of $\cN=2$ supersymmetry are satisfied. 
At this point there are however two subtleties that do not arise in supersymmetry but also do not arise in $\N=1$ supergravity. 
These two new elements are related to the compensators: 
\begin{itemize} 

\item In $\N=2$ we have {\it two} compensating multiplets: ${W}^0$ and ${\cal G}^{ij}$. 
As a result the form of the new Fayet--Iliopoulos is not uniquely fixed by the superconformal invariance. 

\item The ${W}^0$ compensator is a reduced chiral multiplet and can be chosen to be
 the $W$ multiplet that enters the new FI term. 
 We already commented how in this case the construction of the composite goldstino multiplet might be subtle.
Moreover, the new type of FI term for ${W}^0$ will give rise to gravitino higher-derivative terms that should be treated with care. 
Therefore we will not  consider this possibility further in this work. 
From here on, we will assume
\be
W=W^1~,
\ee
and focus for simplicity on a model for a single physical vector multiplet.

\end{itemize}

Let us then proceed to introduce the new FI term in supergravity that we will consider in this paper. 
A rather natural generalization of \eqref{FIgamma} in curved superspace is\footnote{Similarly to the flat FI terms of
\eqref{moreFIgamma}, the function $\cH$ might also depend on the physical vector multiplet $W$ and $\bar{W}$. For simplicity 
we will not investigate this option in this paper.} 
\be
\label{FIgammaSG}
{\cal L}_{\text{new FI}}  
=
 -  \tilde \xi \,  \int \rd^8 \theta \, E \,|\Gamma|^8  \, \cH\left({|{W}^0|^2}/{\cG}\right) \,
{\cal G}^{ij} \, \nabla_{ij} W  \, + c.c. \, , 
\ee
with $\tilde \xi$ is a complex constant, 
and the function $\cH$ is primary of dimension and $U(1)_R$ charge zero, which is identically satisfied by the requirement
of having the combination $|W^0|^2/\cG$ of the compensators as its argument.
For simplicity, in the following we will consider the simple choice for $\cH(|{W}^0|^2/\cG)$:
\be
\cH(|{W}^0|^2/\cG)=\left[ \frac{{W}^0 \overline{W}^0 }{{\cal G}} \right]^n
~,
\label{HGW}
\ee
with $n$ being a constant integer. 
Moreover, 
for the composite compensator superfields that enter the construction of $\Gamma_{\a i}$ 
via \eqref{primary_gaugino}, \eqref{Xsugra}, and \eqref{Gsugra}
we set 
\be
C = {\cal G} \, , \quad Z = {W}^0 \, .  
\ee 
Notice that as the fermion component field of ${\cal G}_{ij}$ ($\chi_{\alpha i} = \frac13 \nabla_\alpha^j {\cal G}_{ij} |$) 
is eventually set to vanish by  gauge fixing,\footnote{We have not explicitly studied the fermionic sector here, 
but when the auxiliary fields are integrated out, 
the fermions $\chi_i$ can be always consistently gauge fixed to vanish by performing an $S$-susy transformation,
see, e.g., \cite{FVP}.}
then the {\it primary gaugini} defined in \eqref{primary_gaugino} will directly be proportional to the gaugini of  ${W}$. 
We can also relate the supersymmetry transformation \eqref{dLdX} to \eqref{DDLLL} once we 
use the compensators ${\cal G}$ and $W^0$. 
Indeed, 
we have 
\be
\cX_{ij} 
=\left[
G^{-1} X_{ij} 
-4 G^{-1/2} (S_{ij}|) \,\phi \right]  
\, , 
\quad 
S^{ij}| = \frac{1}{4 G^{3/2}} (\nabla^{ij} \cG|)  
=  \frac{G^{ij}}{2 G^{5/2}} \Big(F- \frac{G_{kl}}{2 G^2}\chi^{kl}\Big) \, , 
\ee
which after gauge fixing $G=1$ and $\chi^i=0$ gives 
\be
\cX^{ij}  = X^{ij} - 2 \delta^{ij} F \phi \, .  
\ee 
As we will see the vacuum structure of the theory depends significantly on the way the compensators are introduced and 
in particular the integer $n$. 
Then, for this choice, it is simple to check that, in component form, \eqref{FIgammaSG} is 
(we leave here the compensator $G$ manifest for clarity) 
\be
\begin{aligned}
{\cal L} _{\text{new FI}} 
 = - e \, G^{ij} \left\{ \tilde \xi  \, X_{ij}  + c.c. \right\} \, \frac{(\phi^0 \overline \phi^0)^n}{G^n} + \text{fermions} \, ,  
\end{aligned}
\ee
which leads to linear terms for the auxiliary fields $X_{ij} $ of the multiplet $W$. 
Such linear terms in $X_{ij} $ will lead  to a non-vanishing value for $X_{ij} $ once it is integrated out 
and in turn will guarantee the self-consistency of the construction by giving $\langle {\cal X}_{ij} \rangle \ne 0$. 
As we have explained, 
the latter condition will hold when the vacuum breaks completely supersymmetry 
and a verification for this 
will be provided by simply inspecting if the condition \eqref{susyVAC} is violated. 
Otherwise, 
the new FI term we introduce in \eqref{FIgammaSG} will 
typically become singular if the vevs of the fields ${\cal X}^{ij}$ and ${\cal X}^{ij}{\cal X}_{ij}$ vanish.

Because of the existence of more than a single type of new FI term one can have a scenario where not only a 
single type of FI term is switched on. 
For example, 
we could have 
\be
\label{multimulti}
{\cal L}_{\text{multi FI}}  
=
 -  \sum_n \tilde \xi^{(n)} \,  \int \rd^8 \theta \, E \,|\Gamma|^8  \, \left[ \frac{{W}^0 \overline{W}^0 }{{\cal G}} \right]^n 
{\cal G}^{ij} \, \nabla_{ij} W  \, + c.c. \, , 
\ee
leading to all sorts of effective potentials once the auxiliary fields are integrated out. 
The term \eqref{multimulti} can be also considered as an expansion of the term with 
$\cH(|{W}^0|^2/\cG)$ in powers of $|{W}^0|^2/\cG$. 
We will however mostly focus on only one term as in \eqref{HGW} for the rest of our discussion.

We can now introduce the new FI terms into the models studied in the previous section with the aim to decipher 
whether de Sitter vacua 
generically arise in this setup. 
Let us consider the Lagrangian 
\eqref{standardSG} 
where we also add the new FI term for a single vector multiplet. 
For the vector multiplet that enters the composite $\Gamma$ we will set $W=W^1$ 
and we will consider the Lagrangian 
\be 
\label{NNNFI}
{\cal L} = \left\{ - {\cal L}_{\rm V} + c.c. \right\} 
+ {\cal L}_{\rm L} 
+ {\cal L}^{({W}^1)}_{\text{new FI}}  \, . 
\ee
Notice that there is no standard Fayet--Iliopoulos term introduced, 
so the theory here is {\it ungauged}. 
In a standard supergravity setup this theory would have a 
vanishing scalar potential as we have explained, 
however, 
we will see now that a scalar potential will be introduced because of the new FI term. 
All our discussion on the bosonic sector of ungauged $\N=2$ supergravity will be the same 
giving rise to  \eqref{VST} and \eqref{MAXWELL}, 
except of the part that 
contributes to the scalar potential. 
In particular, 
by integrating out the scalar auxiliary fields we will find\footnote{In the ungauged case with new FI term considered, on-shell
it always hold $F=0$ which implies that $\langle X_{ij}\rangle=\langle \cX_{ij}\rangle$.} 
\be
\label{scalAUX}
X^I_{ij} = - 4 \, N^{IJ} \zeta_J \, \delta_{ij} \, , \quad  F = 0 \, , 
\ee
where $N^{IJ} N_{JK} = \delta^I_K $ and 
\be
\label{shiftFI} 
\zeta_I =  8 \tilde \xi \delta^1_I \, \text{e}^{nK} \, . 
\ee 
Here $K$ is the K\"ahler potential defined in \eqref{KAHLER}. 
For the $R$-symmetry auxiliary fields we find 
\be
\quad \Psi_m{}^{ij} = 0 \, ,
\quad  \hat \phi_a = 0 \, , 
\quad \tilde h_m = 0 \, , 
\ee
and therefore, on-shell, all the $SU(2)_R$ symmetry stops to be gauged. 
Eventually we find the bosonic sector of \eqref{NNNFI} to be given by 
\be
\label{VSTnew}
{\cal L} = 
{\cal L}^{\text{ungauged}}_{\text{scalar-gravity}} 
+ {\cal L}_{\text{Maxwell}} 
- e\, {\cal V}_{\text{new FI}} \, . 
\ee
The scalar potential ${\cal V}$ that enters \eqref{VSTnew} takes the form 
\be
\label{newV}
{\cal V}_{\text{new FI}} = - N^{IJ} \zeta_I \zeta_J  \, .  
\ee
We see that the theory has a non-trivial scalar potential without introducing any gaugings. 
This novel potential generated by the new FI terms in ungauged $\cN=2$ supergravity will 
potentially play an important role if one adds also physical hypermultiplets to the supergravity-matter system. 
Notice that in \eqref{NNNFI} the gravitini will have no Lagrangian mass terms because there is no gauging.

Before we study a specific example let us comment on the properties of the scalar potential \eqref{newV}: 
First, 
we point out that if in addition we include new FI terms for more than one physical multiplets, say ${W}^i$, 
we will find that $\zeta_I = \xi_I + 8 \tilde \xi^{i} \delta^i_I \, \text{e}^{n_i K} $ where the $\tilde \xi^i$ are the real 
FI constants for the new FI terms of each 
physical vector multiplet and $n_i$ the integers that determine how the compensators enter. 
Secondly, 
notice that the way the new FI parameters enter into the scalar potential is simply 
by shifting the parameters of the would-be standard FI terms. 
However, due to the specific form of the new FI terms that we have chosen,
this shift does not appear in all the terms induced by  the standard FI terms.
In particular, from \eqref{scalAUX} and \eqref{shiftFI} we see that the shift happens only for the auxiliary fields of the physical 
vector multiplets 
but not for the compensating vector, neither for the tensor multiplet compensator auxiliary field $F$. 
Finally, when $\tilde \xi^i \equiv 0$ we get the standard $\N=2$ ungauged supergravity.

We now turn to an explicit model for the construction of stable dS vacua with a single physical vector multiplet. 
As discussed in the previous section, such vacua will always have spontaneously broken supersymmetry therefore the self-consistency of our constructions here is guaranteed. 
We will have ${\mathbb C}P^{1}$ target space 
and we will allow $n$ to take generic values such that the impact of $n$ on the vacuum structure is clarified. 
We choose the ${\cal F}$ of \eqref{CCPP11} 
therefore the K\"ahler potential and the K\"ahler metric take the form \eqref{KgCP1}. 
The kinetic terms for the vectors in this example are consistent in any background because 
$\omega_{I J} = - \delta_{IJ}$. 
The scalar potential then takes the form 
\be 
\label{NFIV}
{\cal V}_{\text{new FI} -  {\mathbb C}P^{1}} =  \frac{64 \, \tilde \xi^2 }{(1-|z|^2)^{2n}} \, . 
\ee 
We would like to study the vacuum structure of the scalar potential \eqref{NFIV}.  
There are 3 possibilities: 
\begin{itemize}

\item For $n>0$ the scalar potential will have the form ${\cal V} =  64 \, \tilde \xi^2 + 128 \, \tilde \xi^2 \,  n \, |z|^2 + {\cal O}(|z|^4)$. 
As a result the theory generically has a stable de Sitter critical point at $z=0$.

\item For $n=0$ the scalar potential is a constant 
\be 
{\cal V}(z, \overline z) =  64 \, \tilde \xi^2 \, , 
\ee 
and the theory has a positive vacuum energy with a complex modulus $z$. 
This setup provides the simplest model as it contains only gravitation with a positive cosmological constant, 
two gravitini with vanishing {\it Lagrangian} mass (see \cite{Bergshoeff:2015tra} for a discussion on the gravitino mass in de Sitter), 
a massless complex scalar, and two massless abelian vectors.

\item For $n<0$ the scalar potential has no critical points within the moduli space ($|z|<1$) and it will essentially 
describe backgrounds with runaway behavior, 
which drives the scalar towards the boundary of the moduli space. 

\end{itemize}

Constructing stable de Sitter vacua is not always as straightforward as it is for the ${\mathbb C}P^{1}$ model, 
if we restrict ourselves to a single new FI term. 
For example, another class of models that we can consider are the so called $t^3$ models \cite{Cremmer:1984hj}. 
For these models we have 
\be
{\cal F} = - \ri \frac{(\phi^1)^3}{\phi^0} \, , 
\ee
which for $z=s+\ri t$ gives 
\be
N^{11} = \frac{t^2 -3 s^2}{12 t^3} \, , \quad e^K = \frac{1}{8 t^3} \, . 
\ee 
A known feature of the $t^3$ model is that the standard FI term does not lead to any scalar potential despite the gauging.
On the other hand, the presence of a new FI term induces
a scalar potential for the $t^3$ model of the form
\be 
\label{NFIV-3}
{\cal V}_{\text{new FI}-t^3}=  \frac{16 \, \tilde \xi^2 (3 s^2 - t^2 ) }{3 t^3 (8 t^3)^{2n} } \, .
\ee 
Even though this term evades the {\it no-potential} restriction of the $t^3$ model, due to its destabilising runaway behaviour,
it clearly does not have stable de Sitter vacua for any value of $n$.

Let us also note that once the auxiliary fields $X_{ij} $ have acquired a non-vanishing vev 
the construction of the composite $\Gamma$ goldstino is straightforward 
and therefore there is always the possibility to include in the effective theory 
a pure uplift term of the form 
\be
\label{upup-0} 
{\cal L}_{\text{Uplift}} 
=
 -  \int \rd^8 \theta \, E \,|\Gamma|^8  \, \frac{\left[ {W}^0 \overline{W}^0 \right]^{n+2}}{{\cal G}^n} 
= - e \, (\phi^0 \overline \phi^0)^{n+2}  + \text{fermions} \, . 
\ee
The uplift term \eqref{upup-0} is independent of the gauged/ungauged version of the theory, 
and it can be introduced as long as the $X_{ij} $ have acquired a non-vanishing vev, namely $\langle X^{ij}X_{ij}\rangle\ne0$. 
In the case $n=0$, this is the same structure of Volkov--Akulov type of the positive cosmological constant uplift term that was used
to construct in \cite{Kuzenko:2017zla} the off-shell $\cN=2$ extension of 
pure de Sitter supergravity \cite{Bergshoeff:2015tra,Dudas:2015eha} (see \cite{Deser:1977uq,Lindstrom:1979kq} 
for seminal papers on the $\cN=1$
case).
It is also possible to have an uplift term which is a function of the scalar primaries
\be
\label{upup} 
{\cal L}_{\text{Uplift}} 
=
 -  \int \rd^8 \theta \, E \,|\Gamma|^8  \, U(\cG,{W}^I,\overline{W}^I)
 = - e \, U(G,A^I,\overline A^I)  + \text{fermions} 
\, ,
\ee
and extend \eqref{upliftglobal}. The only constraint on $U(\cG,{W}^I,\overline{W}^I)$ is to have dilatation weight 4 and to be 
uncharged under $U(1)_R$.
It is clear that this uplift term, which we stress is self-consistent only when 
$\langle X^{ij} X_{ij} \rangle\ne0$, can lead to any sort of vacua.
More in general, the assumption  $\langle X^{ij} X_{ij} \rangle\ne0$ and $\langle \Delta X \rangle\ne0$
allows to write terms in the effective action where the uplift function $U(\cG,{W}^I,\overline{W}^I)$ is modified to any function
which is a primary of weight 4 also dependent on
$\cD^{ij}\cW$,  $\cD_a\cW$, $\cD_{(\a}^k\cD_{\b)k}\cW$, 
and $S^{ij}$, $X^{\a\b}$, $G_\b{}^\ad$, $G_\b{}^\ad{}_i^j$, $W^{\a\b}$ 
together with their complex conjugate and derivatives $\cD_A$.
Assuming the effective action consistently preserves the condition $\langle X_{ij}\rangle\ne0$, 
$\langle\Delta X\rangle\ne 0$,
the dependence upon the composite superfileds might in general also be non analytic in $X^{ij}X_{ij}$
 leading to a very large freedom.\footnote{See \cite{Kuzenko:2019vaw} for extensions along these lines
 of the $\cN=1$ new FI terms.}

\subsection{New Fayet--Iliopoulos terms in gauged supergravity} 

In this section we include the new FI term in the gauged theory and we study the vacuum structure. 
We consider a theory of the form 
\be
\label{totalgaugedA}
{\cal L} = \left\{ - {\cal L}_{\rm V} + c.c. \right\} 
+ {\cal L}_{\rm L} 
+ {\cal L}_{\text{standard FI}} 
+ {\cal L}^{({W}^1)}_{\text{new FI}}  \, . 
\ee
The discussion will follow the one we presented for the standard gauged supergravity, 
however, 
by integrating out the scalar auxiliary fields we will find 
\be 
F = -2 \, \xi_I \overline \phi^I \, , \quad 
X^J_{ij} = - 4 \, N^{IJ} \zeta_I \, \delta_{ij} \, , \quad 
\hat \phi_m = - 2 \xi_I v^I_m \, , 
\ee 
where now 
\be
\label{shiftFI+standard} 
\zeta_I = \xi_I +  8 \tilde \xi \delta^1_I \, \text{e}^{nK} \, . 
\ee
The full bosonic sector of the theory has the form \eqref{VST-2}  with scalar potential given by 
\bea
\label{old+newV}
{\cal V} = - N^{IJ} \zeta_I \zeta_J - 2 | \xi_I \phi^I |^2 \, . 
\eea
For the gravitini masses we have 
\be
m_{3/2}^2 = |\xi_I \phi^I |^2 \, . 
\ee

Let us now focus on the ${\mathbb C}P^{1}$ model,\footnote{One can also study a gauged $t^3$ model with both the old and 
new FI terms, 
but the gauging does not change significantly the discussion we had for the ungauged $t^3$ model in the previous subsection.}
and note that  
if we switch on all the FI parameters with 
$\xi_I =  (\mu, \xi )$ and  $\tilde \xi, \xi, \mu \in \mathbb{R}$, 
then the scalar potential reads 
\be
\label{scalarVex}
{\cal V} = \left( \frac{8 \tilde \xi}{(1- |z|^2 )^n} + \xi  \right)^2 
-\mu^2 
- 2 \frac{|\mu + \xi z|^2}{1 - |z|^2} \, . 
\ee 
Note that the gravitini kinetic and mass terms have exactly the same form as in standard gauged supergravity, 
that is they are given by \eqref{GRAVI}. 
In fact if the full action \eqref{totalgaugedA} is evaluated in the unitary gauge it will match exactly with the action 
presented in \cite{Cremmer:1984hj} for a single physical vector multiplet, 
the only difference being that the scalar potential will have the form \eqref{scalarVex}. 

To illustrate the properties of the scalar potential \eqref{scalarVex} 
we will study two limiting cases depending on the values of the FI constants of the standard FI terms.

The first limiting case is to set for the FI constants to be 
\be
\xi_I =  (\mu, 0 ) \, , \quad  \tilde \xi, \mu \in \mathbb{R} \, , 
\ee
such that the scalar potential takes the form 
\be
{\cal V} = \left( \frac{8 \tilde \xi}{(1- |z|^2 )^n}  \right)^2 
-\mu^2 
- \frac{2 \mu^2}{1 - |z|^2} \, . 
\ee 
If we are interested in de Sitter vacua there are 3 possibilities: 
\begin{itemize}

\item For $n>0$ the scalar is stabilized at $z=0$ while the vacuum energy can be tuned and it is given by 
\be
{\cal V} = 64 \tilde \xi^2 - 3 \mu^2 \, .
\ee
Therefore the cosmological constant is not identified with the supersymmetry breaking scale. 
Indeed, 
the supersymmetry breaking scale is 
\be
{\rm F}_{\rm SUSY} = \sqrt{{\cal V} + 3 m_{3/2}^2} =  \sqrt{ 64 \tilde \xi^2 - 3 \mu^2 + 3 \mu^2} = 8 \tilde \xi \, , 
\ee
and the gravitino mass is 
\be
m_{3/2} = \mu \, . 
\ee 

The mass of the scalar $z$ is 
\be
m_z^2 = 128 \, n \, \tilde \xi^2  - 2 \mu^2 \, , 
\ee
and it can be easily tuned to be positive. 
In particular, for a positive vacuum energy we will require 
\be
64 \tilde \xi^2 > 3 \mu^2 \, , 
\ee
which gives for any positive integer $n$ 
\be
m_z^2 > (6 n -2) \mu^2 > 0 \, . 
\ee

\item For $n=0$ (and $\tilde \xi \ne 0$) the scalar potential has again a critical point at $z=0$. 
If the space is de Sitter then the critical point at $z=0$ is unstable and the theory develops a runaway 
behavior that drives the scalar towards the boundaries of the moduli space.

\item When $n<0$ the critical point at $z=0$ is unstable for a de Sitter background and there is no other critical point within 
the moduli space.

\end{itemize}

The second limiting case is to set for the FI constants to be 
\be
\xi_I =  (0,\xi ) \, , \quad  \tilde \xi, \xi \in \mathbb{R} \, , 
\ee
which brings the scalar potential to the form 
\be
{\cal V} = \left( \frac{8 \tilde \xi}{(1- |z|^2 )^n} + \xi  \right)^2 
- 2 \xi^2  \frac{|z|^2}{1 - |z|^2} \, . 
\ee 
We are interested again in de Sitter vacua, 
therefore there are 3 possibilities: 
\begin{itemize}

\item For $n>0$ the scalar potential has a critical point at 
\be
z_0 = 0 \, , 
\ee 
delivering a positive vacuum energy given by 
\be
{\cal V}\Big|_{z_0} = (8 \tilde \xi + \xi )^2 \, , 
\ee
while the gravitino mass vanishes. 
Notice that setting $8 \tilde \xi =- \xi$ is not allowed, 
because it would lead to a vanishing 
vev for $X_{ij}^{(1)}$ as can be seen from \eqref{shiftFI+standard}, 
thus rendering the whole construction inconsistent due to the $1/(X^{ij}{}^{(1)}X_{ij}^{(1)})$ 
terms appearing in the fermionic sector of the new FI term. 
The mass of the scalar $z$ is 
\be
\label{massDS}
m_z^2 \Big{|}_{z_0} = 128 n \tilde \xi^2 + 16 n \tilde \xi \xi -2 \xi^2  \, , 
\ee
and it can be easily tuned to be positive, thus providing a stable de Sitter. 

The critical point $z_0=0$ is however {\it not} the only possibility for stable de Sitter vacua. 
For example, 
if we set $n=1$ and $\tilde \xi = \alpha \xi$, 
then the scalar potential has a consistent critical point ($\partial {\cal V}/\partial z=0$) at $z_\alpha$ with 
\be
1 - |z_\alpha|^2 = \frac{64 \alpha^2}{1-8\alpha} \, . 
\ee 
Clearly there is a bound on the values of $\alpha$ given by $1 > 8\alpha$ such that $z_\alpha$ lies within the moduli space. 
The condition that the vacuum energy is positive gives 
\be
{\cal V}\Big{|}_{z_\alpha} = \xi^2 \left( 2 + \frac{16 \alpha -1}{64 \alpha^2} \right) > 0 \quad 
\Longrightarrow \quad \alpha > \frac{\sqrt 3 - 1}{16} \, , 
\ee
which is compatible with $1 > 8\alpha$. 
The mass of the complex scalar is positive only for 
\be
m_z^2\Big{|}_{z_\alpha}  > 0 \quad \Longrightarrow \quad \alpha < \frac{\sqrt 5 - 1}{16} \, . 
\ee
We conclude that there is a stable de Sitter critical point for $n=1$ for any choice of the FI parameters within the bound 
\be
\label{BOUNDa}
\frac{\sqrt 3 - 1}{16} < \frac{\tilde \xi}{\xi} < \frac{\sqrt 5 - 1}{16} < \frac{1}{8} \, , 
\ee
delivering again a positive cosmological constant that can be tuned. 

Notice that for the parameter values that the critical point $z_\alpha$ is stable the critical point $z_0$ is unstable. 
Indeed, for $n=1$ and for $\alpha$ given by \eqref{BOUNDa} we see that the mass \eqref{massDS} is always tachyonic.

\item For $n=0$ the scalar potential has no stable de Sitter critical points.

\item When $n<0$ there are still de Sitter critical points at $z_0=0$. 
The mass of the scalar at $z_0$ is given by \eqref{massDS} which can be positive even when $n<0$, 
for large $|n|$, 
by tuning the values of $\tilde \xi$ and $\xi$.

\end{itemize}

As already mentioned above, in our analysis we focused on constructing simple models possessing stable de Sitter vacua.
Of course, once both the old and new FI terms are turned on there are also new possibilities for anti-de Sitter vacua, 
which we have not investigated here.
Though generically new anti-de Sitter vacua will arise, 
one has to be always careful that the propagating states satisfy the appropriate unitarity bounds. 
A more detailed analysis of anti-de Sitter vacua will be considered elsewhere.

\section{Summary and outlook}

New Fayet--Iliopoulos terms have been recently introduced for $\cN=1$ supergravity theories that do
not require the gauging of the $R$-symmetry \cite{Cribiori:2017laj,Kuzenko:2018jlz,Antoniadis:2018cpq,Antoniadis:2018oeh}, 
and have been studied and developed in a series of 
publications 
\cite{Farakos:2018sgq,Aldabergenov:2018nzd,Abe:2018plc,Cribiori:2018dlc,Aldabergenov:2019hvl,Ishikawa:2019pnb,Aldabergenov:2017hvp,Kuzenko:2019vaw}. 
To highlight some interesting aspects of these constructions let us mention that new type of scalar potentials can be introduced 
that lead to new possibilities for inflation in supergravity \cite{Antoniadis:2018cpq,Aldabergenov:2017hvp}, 
but also to new possibilities regarding the vacuum structure \cite{Cribiori:2017laj,Antoniadis:2018oeh}, 
while the matter content of the theory is still described by standard $\cN=1$ supermultiplets, 
including the FI gauge multiplet. 

In this work we have presented new types 
of Fayet-Iliopoulos terms in $\N=2$ global and local supersymmetry, generalising the $\N=1$ constructions. 
For the construction and study of the new FI term, we used the formalism of non-linear supersymmetry and conformal supergravity. 
The main properties of the new FI term are: 
\begin{enumerate}[label=(\Alph*)]
\item Its existence requires $\cN=2$ supersymmetry to be spontaneously broken completely to $\N=0$ by the auxiliary fields of an 
abelian vector multiplet; 
\item Its bosonic part is linear in the auxiliary fields of the vector multiplet, justifying the name FI term and the requirement (A);
\item Its coupling to supergravity does not require gauging of the $R$-symmetry, contrary to the standard FI term;
\item In the unitary gauge of $\N=2$ supergravity, the fermionic part of the new FI term can be put to zero if supersymmetry breaking 
occurs only by a vev of the corresponding vector multiplet auxiliary component, defining the goldstino direction;
\item The coupling to supergravity allows for a non-trivial dependence of the coefficient of the linear term in the vector multiplet 
auxiliary fields on the compensating scalar fields of the supergravity multiplet,
therefore giving rise to a non-trivial potential for the scalar component of the 
vector multiplet.
\end{enumerate}

We analysed in detail the particular case of one vector multiplet coupled to $\N=2$ supergravity and found 
in a simple example that the scalar potential can have a de  Sitter minimum with 
the scalar field fixed dynamically, evading past no-go theorems based on standard $\N=2$ gauged supergravity \cite{Catino:2013syn}. 
One striking property of our construction is that we can have stable de Sitter vacua 
with a gravitino mass and a cosmological constant that can be tuned, 
and this can be achieved solely with the use of a single $\cN=2$ abelian vector multiplet. 
Such construction was not possible until now in $\cN=2$ supergravity (as has been explained for example in \cite{Catino:2013syn}), 
therefore our construction is expected to lead to new model building directions both for late time and for inflationary 
$\N=2$ supergravity cosmology \cite{Ceresole:2014vpa}.

It is  worthwhile to note that in the case of $\N=1$ supergravity coupled to a vector multiplet, the new FI term is unique, under the 
requirement that its bosonic part is linear in the D-auxiliary field, and amounts to a constant uplift of the vacuum energy 
\cite{Cribiori:2017laj}. 
The presence of matter however brings an ambiguity that manifests in the induced scalar potential, allowing in particular to break or 
not K\"ahler invariance, 
or even to introduce a new function of the matter superfields \cite{Cribiori:2017laj,Antoniadis:2018cpq}. 
In the case of $\N=2$ supergravity the ambiguity appears already at the level of coupling with one vector multiplet which contains a 
scalar field component. 
Technically, it appears through an arbitrary dependence on the ratio of the two compensators in the superconformal formalism (vector 
and tensor or hyper), as mentioned in the point (E) above. 

More in general, once one adopts the assumption that $\cN=2$ supersymmetry is spontaneously broken, 
as already commented in section \ref{section-new-FI}, a vast freedom of new manifestly supersymmetric terms can be consistently
added to general $\cN=2$ supergravity-matter systems.  
A way to underline the new options available is to look at what are probably the two simplest differences between the standard 
and the new FI terms:
\begin{itemize}
\item[{i)}]
Recall that in the off-shell $\cN=2$ supergravity formulation with a vector and tensor compensators
the bosonic sector of the standard FI term  \eqref{bos-standard-FI}  for a physical vector multiplet
 is governed by a single coupling constant $\xi$ and includes three different terms
\be
e^{-1} {\cal L}_{\text{standard FI}} 
= 
 \xi \Big\{
 -\frac18 G^{ij} X_{ij}
- F \phi
+ \frac14 \ve^{mnpq} b_{mn} F_{pq}  
\Big\}
+ c.c. 
= 
\xi\Big\{
-\frac18G^{ij} X_{ij}
- F \phi
+ \tilde{h}^av_a
\Big\}
+ c.c. 
\label{standardFIconclusion}
\ee
As we have reviewed in details in section \ref{review-off-shell-sugra}, the $b_2\wedge F_2$ coupling 
is responsible for the gauging of the $\widehat{U}(1)_R\subset SU(2)_R$ $R$-symmetry, while the second term is the one 
responsible to introducing the universally negative contribution $- 2 | \xi \phi |^2$ to the scalar potential, see eq.~\eqref{oldV}.
Once we assume that local $\cN=2$
supersymmetry is spontaneously broken by the vector multiplet, and then the composite goldstino multiplet 
defined by $X$ and $\G_{\a i}$
 in section \ref{composite-goldstino} is well defined, one has 
the freedom to take apart each of the three terms in 
\eqref{standardFIconclusion} and introduce a supersymmetric Lagrangian of the form
(we neglect for simplicity the possible 
dependence on extra functions of the compensators)
\be
e\Big\{\tilde \xi  \, G^{ij} X_{ij}  
+\tilde \z \,F \phi
+\tilde \rho\, \ve^{mnpq} b_{mn} F_{pq}  
\Big\}
+ c.c. 
 + \text{fermions} \, . 
 \label{newFIconclusion}
\ee
This is parametrized by three arbitrary constants $\tilde\x$, $\tilde{\z}$, and $\tilde\rho$ and,
by playing with this new freedom, one can tune the different
physical consequences that each bosonic term has. 
The first term is typically necessary to be there since it is the one dominating the condition $\langle \D X\rangle\ne0 $.
For this reason and for simplicity in the paper we focused on the new FI term where
$\tilde\x\ne0$ and $\tilde{\z}=\tilde\rho=0$ and  considered as an extension a linear combination of standard and new FI terms.

\item[ii)]
Another simple difference that spontaneously broken $\cN=2$ supersymmetry allows 
(by using non-linear realization techniques) is the possibility to have 
uplift terms of the form \eqref{upup} governed by an a priori arbitrary function  of the primary scalar fields in 
the theory, analogously to the \emph{liberated} $\cN=1$ supergravity of \cite{Farakos:2018sgq}. 
\end{itemize}
In this paper we have focused on the case where supersymmetry is broken by a single vector multiplet,
but similar analysis can  straightforwardly be performed when supersymmetry breaking is mediated by more than one 
physical vector multiplet 
and/or other multiplets, as for instance systems of hypermultiplets. 
These constructions will naturally overcome known no-go theorems as, for instance, the impossibility to 
introduce standard FI terms whenever physical charged hypermultiplets are coupled to $\cN=2$ supergravity
\cite{DAuria:1990qxt,Andrianopoli:1996vr,Andrianopoli:1996cm,DallAgata:2003sjo,VanProeyen:2004xt,FVP,Trigiante:2016mnt,Bagger:1987rc}. 
New FI terms are a natural option to overcome the constraints on the couplings with charged hypermultiplets
that come with the gauging of isometries in the quaternionic-K\"ahler geometry.
Here we have only scratched the surface of the effective $\cN=2$ supergravity theories that can be constructed 
using the ideas in our paper.

Another aspect that will deserve further studies is the choice of off-shell Poincar\'e supergravity one starts from.
In our paper 
we have chosen a description given by $\cN=2$ conformal supergravity coupled to a vector and a tensor multiplet compensators
\cite{sct_structure,deWPV} which can be considered as an $\cN=2$ analogue of the  new-minimal formulation of 4D $\cN=1$ 
off-shell supergravity (see \cite{GGRS,Wess:1992cp,FVP,Ideas} 
for reviews of the different off-shell formulations of $\cN=1$ Poincar\'e supergravity). 
While the vector multiplet represents a standard choice of compensator for $\cN=2$ off-shell 
supergravity since it fixes $U(1)_R\subset U(2)_R$, 
the choice of the tensor multiplet cannot fix the $SU(2)_R$ factor
 leaving a residual $\widehat{U}(1)_{R}$  symmetry off-shell (that, depending on the model, is eventually broken on-shell).
 This restrict the classes of matter theories that can be coupled to the off-shell Poincar\'e supergravity
 that we have employed in our work.
Variant choices of  the hypermultiplet compensator, such as the scalar multiplet or the non-linear multiplet originally used in 
\cite{sct_structure}, allow to completely fix the $SU(2)_R$. 
Alternatively, one could use an off-shell hypermultiplet compensator 
that, without central charges, 
is known to lead in general to an infinite set of auxiliary fields that can efficiently be handled by using harmonic 
\cite{GIKOS,GIOS,SUGRA-har}
or projective superspace 
techniques \cite{KLR,LR3,LR2}. General 4D $\cN=2$ off-shell supergravity-matter couplings can be described in a covariant way 
by using the superspace techniques of \cite{KLRT-M,Kuzenko:2009zu}.\footnote{See also 
\cite{Projective-Dan-1,Projective-Dan-2,Butter4DN=2,Butter:2012xg} 
for further extensions of the formalism
and \cite{ProjectiveSugra5D,ProjectiveSugra2D,ProjectiveSugra3D,LT-M12} 
for curved projective superspace techniques in $D=2,3,5,6$ dimensions.}
By using these approaches, it would be natural to extend the analysis of our work and study new FI terms in general systems of 
off-shell hypermultiplets.

Among the most important questions left open is how to constrain the plethora of models with spontaneously broken $\cN=2$ 
supersymmetry that can be constructed by using the ideas of our paper.
In fact, it would very interesting to see whether there is a possible microscopic origin of a new FI term, for instance in string theory. 
This would be particularly important in view of the recent {\it swampland} conjectures (see e.g. \cite{Palti:2019pca} for a review) 
related to the 
existence or not of de Sitter vacua in quantum (super) gravity theories. 
Consistency arguments constraining the low energy effective field and supergravity theories 
will hopefully give clear criteria on the allowed new terms. 
 
Another natural question is whether there exist possible variations of 
the new FI terms in the case of $\N=2\to\N=1$ partial supersymmetry breaking \cite{Antoniadis:1995vb} 
(for the local supersymmetric case see, for example, \cite{Ferrara:1995xi} and more recently \cite{Antoniadis:2018blk})
or $\N=2\to\N= 0$ at two different scales~\cite{Antoniadis:2012cg}. 
For supersymmetry breaking mediated by a vector multiplet,
both cases require most likely to introduce deformations of its supersymmetry transformations 
corresponding to magnetic-type FI terms \cite{Antoniadis:1995vb}
whose superspace description relevant to extending our analysis can be found in  
\cite{IZ1,IZ2,RT,Kuzenko:2015rfx,Antoniadis:2017jsk,Antoniadis:2019gbd,Cribiori:2018jjh,Kuzenko:2013gva,Kuzenko:2017gsc} 
both for the global and local cases.

\section*{Acknowledgements} 
We thank Stefanos Katmadas, 
Sergei Kuzenko and Antoine Van Proeyen for discussions. 
The work of IA was supported in part by the Swiss National Science Foundation, 
in part by the Labex ``Institut Lagrange de Paris'' and in part by a CNRS PICS grant.
The work of FF is supported from the KU Leuven C1 grant ZKD1118 C16/16/005. 
GT-M is supported by the Albert Einstein Center for Fundamental Physics, University of Bern,
and by the Australian Research Council (ARC) Future Fellowship FT180100353.
FF thanks for support and hospitality the Albert Einstein Center for Fundamental Physics, University of Bern, 
during the early stages of this work.

\appendix

\section{Nilpotent chirals from generic $\cN=2$ supersymmetry breaking} 
\label{Proof-Nilpotency}

In this section we show that when $\cN=2$ supersymmetry is broken one can always construct an $\cN=2$ 
nilpotent chiral multiplet that will describe the goldstini. 
We will use the results of this appendix in the main bulk of the article focusing on the vector multiplet. 
Let us assume that $\cN=2$ supersymmetry is broken spontaneously and the goldstini of this 
$\cN=2\to\cN=0$ breaking are the 
lowest components of the $\cN=2$ superfields $\Psi_\alpha$ and $\Xi_\alpha$, that, in $SU(2)_R$ notations, can be 
collected in an $SU(2)_R$ doublet complex spinor superfield $\Psi_\a^i=(\Psi_\a,\Xi_\a)$. 
These two superfields may be constrained, as it happens in the vector multiplet, 
but our analysis here holds for a generic setup where it is not necessary to specify the conditions satisfied by $\Psi_\a^i$. 
First we observe a series of nilpotency conditions that hold even for superfields that do not include the goldstino. 
These nilpotency conditions are only satisfied because of the large number of fermions. 
We define the $\beta$ that is the maximum product of goldstini to be  
($\Psi^{ij}:=\Psi^{\a i}\Psi_\a^j$, $\Psi^4:=\frac{1}{3}\Psi^{ij}\Psi_{ij}$, $|\Psi|^8=\Psi^4\overline\Psi^4$)
\be
\beta = \Psi^2 \Xi^2 \overline{\Psi}^2 \overline{\Xi}^2
=|\Psi|^8 \,. 
\ee
Clearly we can see that 
\be
\beta^2 =0 \, , 
\ee
but it also satisfies a series of nilpotency conditions of the form 
\be
\left(D_{A_1}D_{A_2}\cdots D_{A_m} \beta\right) 
D_{B_1}D_{B_2}\cdots D_{B_n}\beta ~ =0
~,~~~~~~
\forall~
m,n=0,1,\cdots,7~,~~~
m+n\leq 7
~, 
\label{nilpotencies}
\ee 
where $D_A$ refers collectively to $(\partial_a,D_\alpha, \overline D^{\dot \alpha},\tilde D_\alpha, \overline{\tilde D}^{\dot \alpha})$
or $(\partial_a,D^i_\alpha, \overline D^{\dot \alpha}_i)$
in $SU(2)_R$ notations. 
By introducing the chiral projector operator $\overline{\D}:=\frac{1}{48}\overline D^4$ with 
$\overline D^4:=\DB_\ad^k\DB^{\ad l}\DB_{\bd k}\DB^{\bd}_{l}$ 
as in eq.~\eqref{chiral-projector-flat},
we can construct an $\cN=2$ chiral nilpotent superfield 
\be
X = \overline \D^4  \beta \, , 
\ee
that satisfies 
\be
\label{NILone}
X^2 = 0
 \, , \quad 
X D_A X = 0
 \, , \quad 
X D_AD_B X = 0
 \, , \quad 
X D_AD_BD_C X = 0
 \, . 
\ee
To prove the first property in \eqref{NILone} we observe that 
\be
X^2 =(\overline \D^4 \beta) (\overline \D^4 \beta)  
= \overline \D^4\left( \beta \ \overline \D^4\beta \right)  
= 0 \, . 
\ee
The rest of the properties in \eqref{NILone}  
are derived in a similar manner, 
and they reduce to identities of the form 
$\overline D^4 \left( \beta D_{A}\overline{D}^4\beta \right)=0$,
$\overline D^4 \left( \beta D_{A}D_{B}\overline{D}^4\beta \right)=0$,
and
$\overline D^4 \left( \beta D_{A}D_{B}D_{C}\overline{D}^4 \beta \right)=0$
that are identically satisfied due to \eqref{nilpotencies}.
To conclude, we stress that the above construction of a composite nilpotent chiral multiplet works for a completely 
arbitrary spinor superfield $\Psi_\a^i$ and could be used in principle starting with multiplets other than the vector one.
The only extra necessary condition required to construct  a composite goldstino multiplet for 
$\cN=2\to\cN=0$ supersymmetry breaking mediated by $\Psi_\a^i$ is that 
$\langle D^4X\rangle=\langle D^4\overline D^4\b\rangle\ne0$.

\section{$\cN=2$ conformal superspace} 
\label{AppendixConfSuperspace}

This appendix contains a summary of the formulation for $\cN=2$ conformal supergravity  
in conformal  superspace\footnote{Conformal superspace was 
 first introduced by D.~Butter for 4D $\cN=1$ \cite{Butter:2009cp} 
and $\cN=2$  \cite{Butter4DN=2} supergravity
(see also the seminal work by Kugo and Uehara \cite{Kugo:1983mv} and the recent paper \cite{Kugo:2016zzf})
and it was developed and extended to 3D $\cN-$extended supergravity \cite{Butter:2013goa}, 
5D $\cN=1$ supergravity \cite{Butter:2014xxa},
and recently to 6D $\cN=(1,0)$ supergravity \cite{Butter:2016qkx},
see also \cite{Butter:2017jqu}.
}
 \cite{Butter4DN=2} employed in sections \ref{review-off-shell-sugra} and \ref{NewFI-sugra}.
We use the notations of \cite{Butter:2012xg} 
and review the results necessary for deriving results in sections \ref{review-off-shell-sugra} and \ref{NewFI-sugra}.
The structure group of $\cN=2$ conformal superspace 
is chosen to be $SU(2,2|2)$
and the covariant derivatives  
$\nabla_A = (\nabla_a, \nabla_\a^i , \overline{\nabla}^\ad_i)$ have the form
\begin{align} \nabla_A &= E_A + \hf \Omega_A{}^{ab} M_{ab} + \Phi_A{}^{ij} J_{ij} + \ri \Phi_A Y 
+ B_A \mathbb{D} + \frak{F}_{A}{}^B K_B \non\\
&= E_A + \Omega_A{}^{\b\g} M_{\b\g} + \overline{\Omega}_A{}^{\bd\gd} \overline{M}_{\bd\gd} + \Phi_A{}^{ij} J_{ij} + \ri \Phi_A Y 
+ B_A \mathbb{D} + \frak{F}_{A}{}^B K_B \ .
\end{align}
Here, 
$E_A =E_A{}^M \pa_M$ is the inverse of the supervielbein super one-form 
$E^A=\rd z^M E_M{}^A$, $M_{cd}$ and $J_{kl}$ are the generators of the Lorentz and $SU(2)_R$ $R$-symmetry groups 
respectively, and $\O_A{}^{bc}$ and $\Phi_A{}^{kl}$ the corresponding connections. 
The remaining generators and corresponding connections are: 
$Y$  and $\Phi_A$ for the $U(1)_R$ $R$-symmetry group; 
$\mathbb D$ and $B_A$ for the dilatations;
 $K^A = (K^a, S^\a_i, \overline{S}_\ad^i)$
and $\frak{F}_A{}^B$
for the special superconformal generators.

The Lorentz and $SU(2)_R$ generators act on $\de_A$ as 
\bea
[M_{\a\b},  \de_\g^i] = \ve_{\g (\a } \de_{\b)}^i 
\ , \qquad 
\big[ J_{kl}, \de_\a^i \big] = - \d^i_{(k} \de_{\a l)} 
\ ,
\label{Lorentz-SU(2)-Gen}
\eea
together with their complex conjugates.
The $U(1)_R$ and dilatation generators obey
\bsubeq
\bea
&{[}Y, \nabla_\a^i{]}= \nabla_\a^i ~,\quad {[}Y, \overline\nabla^\ad_i{]} = - \overline\nabla^\ad_i~,
\\
&{[}\mathbb{D}, \nabla_a{]} = \nabla_a ~,\quad
{[}\mathbb{D}, \nabla_\a^i{]} = \hf \nabla_\a^i ~,\quad
{[}\mathbb{D}, \overline\nabla^\ad_i{]} = \hf \overline\nabla^\ad_i ~.
\eea
\esubeq
The special superconformal generators $K^A$ transform 
under  Lorentz and $SU(2)_R$  as
\bea
&[M_{ab}, K_c] = 2 \eta_{c [a} K_{b]} ~, \quad
[M_{\a\b} , S^\g_i] =\d^\g_{(\a}S_{\b)i} ~, \quad
[J_{ij}, S^\g_k] = - \ve_{k (i} S^\g_{j)} ~,
\eea
together with their complex conjugates,
while their transformation under $U(1)_R$ and dilatations is given by:
\bsubeq
\bea
&[Y, S^\a_i] = - S^\a_i ~, \quad
[Y, \overline{S}^i_\ad] = \overline{S}^i_\ad~, \non \\
&[\mathbb{D}, K_a] = - K_a ~, \quad
[\mathbb{D}, S^\a_i] = - \hf S^\a_i ~, \quad
[\mathbb{D}, \overline{S}_\ad^i] = - \hf \overline{S}_\ad^i ~.
\eea
\esubeq
The generators $K^A$ obey
\begin{align}
\{ S^\a_i , \overline{S}^j_\ad \} &= 2 \ri \d^j_i (\s^a)^\a{}_\ad K_a
~,
\end{align}
while the nontrivial (anti-)commutators of the algebra of $K^A$ with $\nabla_B$ are given by
\bsubeq
\bea
&[K^a, \nabla_b] = 2 \delta^a_b \mathbb{D} + 2 M^{a}{}_b 
~,\non \\
&\{ S^\a_i , \nabla_\b^j \} = 
2 \d^j_i \d^\a_\b \mathbb{D} - 4 \d^j_i M^\a{}_\b 
- \d^j_i \d^\a_\b Y + 4 \d^\a_\b J_i{}^j ~,\non \\
&[K^a, \nabla_\b^j] = -\ri (\s^a)_\b{}^\bd \overline{S}_\bd^j \ , \quad
[S^\a_i , \nabla_b] = \ri (\s_b)^\a{}_\bd \overline{\nabla}^\bd_i 
\ ,
\eea
\esubeq
together with complex conjugates.

The  (anti-)commutation relations of the covariant derivatives $\de_A$ \cite{Butter4DN=2,Butter:2012xg}
relevant for calculations in this paper are
\begin{subequations}\label{CSGAlgebra}
\begin{align}
\{ \nabla_\a^i , \nabla_\b^j \} &= 2 \ve^{ij} \ve_{\a\b} \overline{W}_{\gd\dd} \overline{M}^{\gd\dd} 
+ \hf \ve^{ij} \ve_{\a\b} \overline{\nabla}_{\gd k} \overline{W}^{\gd\dd} \overline{S}^k_\dd 
- \hf \ve^{ij} \ve_{\a\b} \nabla_{\g\dd} \overline{W}^\dd{}_\gd K^{\g \gd}~, \\
\{ \nabla_\a^i , \overline{\nabla}^\bd_j \} &= - 2 \ri \d_j^i \nabla_\a{}^\bd~, \\
[\nabla_{\a\ad} , \nabla_\b^i ] &= - \ri \ve_{\a\b} \overline{W}_{\ad\bd} \overline{\nabla}^{\bd i} 
- \frac{\ri}{2} \ve_{\a\b} \overline{\nabla}^{\bd i} \overline{W}_{\ad\bd} \mathbb{D} 
- \frac{\ri}{4} \ve_{\a\b} \overline{\nabla}^{\bd i} \overline{W}_{\ad\bd} Y 
+ \ri \ve_{\a\b} \overline{\nabla}^\bd_j \overline{W}_{\ad\bd} J^{ij}
	\eol & \quad
	- \ri \ve_{\a\b} \overline{\nabla}_\bd^i \overline{W}_{\gd\ad} \overline{M}^{\bd \gd} 
	- \frac{\ri}{4} \ve_{\a\b} \overline{\nabla}_\ad^i \overline{\nabla}^\bd_k \overline{W}_{\bd\gd} \overline{S}^{\gd k} 
	+ \frac{1}{2} \ve_{\a\b} \nabla^{\g \bd} \overline{W}_{\ad\bd} S^i_\g
	\eol & \quad
	+ \frac{\ri}{4} \ve_{\a\b} \overline{\nabla}_\ad^i \nabla^\g{}_\gd \overline{W}^{\gd \bd} K_{\g \bd}~,
\end{align}
\end{subequations}
together with complex conjugates.
The superfield $W_{\a\b} = W_{\b\a}$, and its complex conjugate
${\overline{W}}_{\ad \bd} := \overline{W_{\a\b}}$, are dimension one conformal primaries, that is 
$K_A W_{\a\b} = 0$, and obey the additional constraints
\bsubeq
\bea
&YW_{\a\b}=-2W_{\a\b}
~,\qquad
Y\overline{W}_{\ad\bd}
=2\overline{W}_{\ad\bd}
~,
\\
&\overline{\nabla}^\ad_i W_{\b\g} = 0~,\qquad
\nabla_{\a}^k \nabla_{\b k} W^{\a\b} =\overline\nabla^{\ad}_k \overline\nabla^{\bd k}  \overline{W}_{\ad\bd} ~.
\eea
\esubeq
The superfield $W_{\a\b}$ is the $\cN=2$ super-Weyl tensor. 
It can be proven that the previous construction describes a superfield embedding of the 
standard Weyl multiplet of $\cN=2$ conformal supergravity. 
See \cite{Butter4DN=2,Butter:2012xg} for details.

In section \ref{NewFI-sugra} we also used the covariant superspace derivatives $\cD_\a^i$ and $\cDB^\ad_i$
 defined in eq.~\eqref{Dcov} 
that are useful to construct primary extensions of multiplets \cite{Butter4DN=2}.
When acting on a conformally primary dimensionless tensor, the algebra of 
these covariant derivatives  becomes 
\begin{subequations}\label{algebraCalD}
\bea
\{\cD_\alpha^i, \cD_\beta^j\}
	&=&
	4 S^{ij} M_{\alpha \beta}
	+ 2 \eps^{ij} \eps_{\alpha \beta} X^{\gamma \delta} M_{\gamma\delta}
	+ 2 \eps^{ij} \eps_{\alpha \beta} \overline W'_{\dgamma \ddelta} \overline{M}^{\dgamma \ddelta}
\non\\
	&&
	+ 2 \eps^{ij} \eps_{\alpha \beta} S^{kl} J_{kl}
	+ 4 X_{\alpha \beta} J^{ij}~,
	 \\
\{\overline\cD^\ad_i, \overline\cD^\dbeta_j\}
	&=&
	-4 \overline S_{ij} \overline M^{\ad \dbeta}
	- 2 \eps_{ij} \eps^{\ad \dbeta} \overline X_{\dgamma \ddelta} \overline{M}^{\dgamma\ddelta}
	- 2 \eps_{ij} \eps^{\ad \dbeta} W'{}^{\gamma \delta} M_{\gamma \delta}
\non\\
&&
	- 2 \eps_{ij} \eps^{\ad \dbeta} \overline S^{kl} J_{kl}
	- 4 \overline X^{\ad \dbeta} J_{ij}~,
\\
\{\cD_\alpha^i, \overline \cD^\ad_j\} &=&
	-2\ri \delta^i_j  \cD_\alpha{}^\ad
	- 2 (G_\alpha{}^\ad \delta^i_j + \ri G_\alpha{}^\ad{}^i{}_j) Y
\non\\
&&
	+ 4 (G_{\alpha \dbeta} \delta^i_j + \ri G_{\alpha \dbeta}{}^i{}_j) \overline M^{\dbeta \ad}
	+ 4 (G^{\ad \beta} \delta^i{}_j + \ri G^{\ad \beta}{}^i{}_j) M_{\beta \alpha}
\non\\
&&
	+ 8 G_\alpha{}^\ad J^i{}_j
	+ 4 \ri \delta^i_j G_\alpha{}^\ad{}^k{}_l J^l{}_k ~, 
\eea
\end{subequations}
with the vector covariant derivative operator 
$\cD_{\alpha \ad} = (\s^a)_{\a\ad} \cD_a$ 
given by
\begin{align}
 \cD_\alpha{}^\ad &:=
	C^{-1/2} \nabla_\alpha{}^\ad
	- \frac{\ri}{2} C^{-1/4} \nabla_\alpha^k U \overline\cD^\ad_k
	- \frac{\ri}{2} C^{-1/4} \overline \nabla^\ad_k U \cD_\alpha^k
 	\eol & \quad
	- \left(\frac{\ri}{4} C^{-1/2} \overline\nabla^\ad_k \nabla^{\beta k} U + 2 \ri G^{\ad \beta} \right)
		M_{\beta \alpha}
	+ \left(\frac{\ri}{4} C^{-1/2} \nabla_\alpha^k \overline\nabla_{\dbeta k} U - 2 \ri G_{\alpha \dbeta} \right)
		\overline M^{\dbeta \ad}
	\eol & \quad
	- \ri \left(\frac{1}{16} C^{-1/2} [\nabla_\alpha^k, \overline\nabla^\ad_k] U - G_\alpha{}^\ad \right) Y
	+ \frac{\ri}{2} C^{-1/2} \nabla_\alpha^k U \overline\nabla^\ad_{j} U  J^j{}_k~,
\end{align}
and the primary dimension zero (they are all invariant under dilatations) curvature superfields 
\begin{subequations}\label{dim-1-torsions}
\begin{alignat}{2}
S^{ij} &:= \frac{1}{4C^{3/2}} \nabla^{ij} C~, &\qquad
\overline S_{ij} &:= \frac{1}{4C^{3/2}} \overline \nabla_{ij} C~, \\
X_{\alpha \beta} &:= -\frac{C^{1/2}}{4} \nabla_{\alpha \beta} C^{-1}~, &
\overline X_{\ad \dbeta} &:= -\frac{C^{1/2}}{4} \overline\nabla_{\ad \dbeta} C^{-1}~, \\
W'_{\a\b} &:= C^{-1/2} W_{\a\b}~,&
\overline W'_{\ad\bd} &:= C^{-1/2} \overline W_{\ad\bd}
~,
\\
G_{\alpha \ad} &:= -\frac{1}{16} C^{1/2} [\nabla_\alpha^k, \overline\nabla_{\ad k}] C^{-1}~,&\quad\quad\quad
G_{\alpha \ad}{}^{ij} &:= -\frac{\ri}{8} C^{-1/2} [\nabla_\alpha^{(i}, \overline\nabla_\ad^{j)}] U
~.
\end{alignat}
\end{subequations}
As explained in \cite{Butter4DN=2,Butter:2012xg},
the geometry described by the previous algebra for the $\cD_A=(\cD_a,\cD_\a^i,\cDB^\ad_i)$ derivatives
 is equivalent to the one introduced by P.~Howe in 1980
\cite{Howe:1980sy}
to describe conformal supergravity in a superspace equipped with a $Sl(2,\mathbb C)\times U_R(2)$ structure group 
(see also \cite{Kuzenko:2009zu}).


\begin{small}

\end{small}

\end{document}